\documentclass[12pt,preprint]{aastex}

\begin{document}

\title{Exploring the Variable Sky with the Sloan Digital Sky Survey}

\author{
Branimir Sesar\altaffilmark{\ref{Washington}},
\v{Z}eljko Ivezi\'{c}\altaffilmark{\ref{Washington}},
Robert H.~Lupton\altaffilmark{\ref{Princeton}},
Mario Juri\'{c}\altaffilmark{\ref{IAS}},
James E.~Gunn\altaffilmark{\ref{Princeton}},
Gillian R.~Knapp\altaffilmark{\ref{Princeton}}, 
Nathan De Lee\altaffilmark{\ref{MSU}},
J.~Allyn Smith\altaffilmark{\ref{AustinPeay}},
Gajus Miknaitis\altaffilmark{\ref{FNAL}},
Huan Lin\altaffilmark{\ref{FNAL}},
Douglas Tucker\altaffilmark{\ref{FNAL}},
Mamoru Doi\altaffilmark{\ref{UT}},
Masayuki Tanaka\altaffilmark{\ref{UT2}},
Masataka Fukugita\altaffilmark{\ref{UT3}},
Jon Holtzman\altaffilmark{\ref{NMSU}},
Steve Kent\altaffilmark{\ref{FNAL}},
Brian Yanny\altaffilmark{\ref{FNAL}},
David Schlegel\altaffilmark{\ref{LBNL}}, 
Douglas Finkbeiner\altaffilmark{\ref{Harvard}},
Nikhil Padmanabhan\altaffilmark{\ref{LBNL}},
Constance M.~Rockosi\altaffilmark{\ref{UCSC}}, 
Nicholas Bond\altaffilmark{\ref{Princeton}},
Brian Lee\altaffilmark{\ref{LBNL}},
Chris Stoughton\altaffilmark{\ref{FNAL}}, 
Sebastian Jester\altaffilmark{\ref{Southampton}}, 
Hugh Harris\altaffilmark{\ref{USNOFlagstaff}}, 
Paul Harding\altaffilmark{\ref{CWRU}}, 
Jon Brinkmann\altaffilmark{\ref{APO}}, 
Donald P. Schneider\altaffilmark{\ref{PennState}},
Donald York\altaffilmark{\ref{Chicago}},
Michael W. Richmond\altaffilmark{\ref{Rochester}},
Daniel Vanden Berk\altaffilmark{\ref{PennState}}
}

\altaffiltext{1}{University of Washington, Dept.~of Astronomy, Box
                           351580, Seattle, WA 98195-1580\label{Washington}}
\altaffiltext{2}{Princeton University Observatory, Princeton,
                           NJ 08544-1001\label{Princeton}}
\altaffiltext{3}{Institute for Advanced Study, 1 Einstein Drive,
                           Princeton, NJ 08540\label{IAS}}
\altaffiltext{4}{Dept.~of Physics \& Astronomy, Michigan State
                           University, East Lansing, MI 48824-2320\label{MSU}}
\altaffiltext{5}{Dept.~of Physics \& Astronomy, Austin Peay State
                           University, Box 4608, Clarksville, TN 37044
                           \label{AustinPeay}}
\altaffiltext{6}{Fermi National Accelerator Laboratory, Box 500,
                           Batavia, IL 60510\label{FNAL}}
\altaffiltext{7}{Institute of Astronomy, University of Tokyo,
                           2-21-1 Osawa,Mitaka, Tokyo 181-0015, Japan\label{UT}}
\altaffiltext{8}{Dept. of Astronomy, Graduate School of Science,
                           University of Tokyo, Hongo 7-3-1, Bunkyo-ku, Tokyo,
                           113-0033, Japan\label{UT2}}
\altaffiltext{9}{Institute for Cosmic Ray Research, University of
                           Tokyo, Kashiwa, Chiba, Japan\label{UT3}}
\altaffiltext{10}{New Mexico State University, Box 30001, 1320 Frenger
                           St., Las Cruces, NM 88003\label{NMSU}}
\altaffiltext{10}{Lawrence Berkeley National Laboratory, One Cyclotron
                           Road, MS 50R5032, Berkeley, CA, 94720\label{LBNL}}
\altaffiltext{11}{Harvard-Smithsonian Center for Astrophysics, 60
                           Garden Street, Cambridge, MA 02138\label{Harvard}}
\altaffiltext{12}{University of California--Santa Cruz, 1156 High St.,
                           Santa Cruz, CA 95060\label{UCSC}}
\altaffiltext{13}{School of Physics and Astronomy, University of
                           Southampton, Highfield, Southampton, SO17 1BJ, UK
                           \label{Southampton}}
\altaffiltext{14}{U.S. Naval Observatory, Flagstaff Station, Box 1149,
                           Flagstaff, AZ 86002\label{USNOFlagstaff}}
\altaffiltext{15}{Department of Astronomy, Case Western Reserve
                           University, Cleveland, Ohio 44106\label{CWRU}}
\altaffiltext{16}{Apache Point Observatory, 2001 Apache Point Road,
                           Box 59, Sunspot, NM 88349-0059\label{APO}}
\altaffiltext{17}{Department of Astronomy and Astrophysics,
                           Pennsylvania State University, University Park,
                           PA 16802\label{PennState}}
\altaffiltext{18}{University of Chicago, Astronomy \& Astrophysics
                           Center, 5640 S. Ellis Ave., Chicago, IL 60637
                           \label{Chicago}}
\altaffiltext{19}{Rochester Institute of Technology, Department of Physics,
                           84 Lomb Memorial Dr., Rochester, NY 14623-5603
                           \label{Rochester}}

\begin{abstract}
We quantify the variability of faint unresolved optical sources using a catalog
based on multiple SDSS imaging observations. The catalog covers SDSS Stripe 82,
which lies along the celestial equator in the Southern Galactic Hemisphere
(22h 24m~$< \alpha_{J2000} <$~04h 08m, $-1.27\arcdeg<\delta_{J2000} <
+1.27\arcdeg$, $\sim290$ deg$^2$), and contains 58 million photometric
observations in the SDSS $ugriz$ system for 1.4 million unresolved sources that
were observed at least 4 times in each of the $gri$ bands (with a median of 10
observations obtained over $\sim$5 years). In each photometric bandpass we
compute various low-order lightcurve statistics such as root-mean-square scatter
(rms), $\chi^2$ per degree of freedom, skewness, minimum and maximum magnitude,
and use them to select and study variable sources. We find that $2\%$ of
unresolved optical sources brighter than $g=20.5$ appear variable at the 0.05
mag level (rms) simultaneously in the $g$ and $r$ bands. The majority (2/3) of
these variable sources are low-redshift ($<2$) quasars, although they represent
only $2\%$ of all sources in the adopted flux-limited sample. We find that at
least $90\%$ of quasars are variable at the 0.03 mag level (rms) and confirm
that variability is as good a method for finding low-redshift quasars as is the
UV excess color selection (at high Galactic latitudes). We analyze the
distribution of lightcurve skewness for quasars and find that is centered on
zero. We find that about 1/4 of the variable stars are RR Lyrae stars, and that
only $0.5\%$ of stars from the main stellar locus are variable at the 0.05 mag
level. The distribution of lightcurve skewness in the $g-r$ vs.~$u-g$
color-color diagram on the main stellar locus is found to be bimodal (with one
mode consistent with Algol-like behavior). Using over six hundred RR Lyrae
stars, we demonstrate rich halo substructure out to distances of 100 kpc. We
extrapolate these results to expected performance by the Large Synoptic Survey
Telescope and estimate that it will obtain well-sampled $2\%$ accurate,
multi-color lightcurves for $\sim2$ million low-redshift quasars, and will
discover at least 50 million variable stars.

\end{abstract}


\section{Introduction}

Variability is an important phenomenon in astrophysical studies of structure and
evolution, both stellar and galactic. Some variable stars, such as RR Lyrae,
are an excellent tool for studying the Galaxy. Being nearly standard candles
(thus making distance determination relatively straightforward) and being
intrinsically bright, they are a particularly suitable tracer of Galactic
structure. In extragalactic astronomy, the optical continuum variability of
quasars is utilized as an efficient method for their discovery
\citep{bhp73,haw83,kkc86,hv95}, and is also frequently used to constrain the
origin of their emission \citep{kaw98,tre01,ms03}.

Despite the importance of variability, the variable optical sky remains largely
unexplored and poorly quantified, especially at the faint end. To what degree
different variable populations contribute to the overall variability, how they
are distributed in magnitude and color, what the characteristic time-scales and
the dominant mechanisms of variability are, are just some of the questions that
still remain to be answered. To address these questions, several contemporary
projects aimed at regular monitoring of the optical sky were started. Some of
the more prominent surveys in terms of the sky coverage, depth, and cadence are:

\begin{itemize}
\item
The Faint Sky Variability Survey \citep{gro03}
is a very deep ($17<V<24$) $BVI$ survey of 23 deg$^2$ of sky, containing about
80,000 sources sampled at timescales ranging from minutes to years.
\item
The QUEST Survey \citep{viv01} monitors 700 deg$^2$ of sky from $V=13.5$ to a
limit of $V=21$.
\item
ROTSE-I \citep{ake00} monitors the entire observable sky twice a night from 
$V=10$ to a limit of $V=15.5$. The Northern Sky Variability Survey \citep{woz04}
is based on ROTSE-I data.
\item
OGLE (most recently OGLE III; \citealt{uda02}) monitors $\sim100$ deg$^2$
towards the Galactic bulge from $I=11.5$ to a limit of $I=20$. Due to the very
high stellar density towards the bulge, OGLE II has detected about 270,000
variable stars \citep{woz02,zeb02}.
\item
The MACHO Project monitored the brightness of $\sim60$ million stars in $\sim90$
deg$^2$ of sky toward the Magellanic Clouds and the Galactic bulge for $\sim7$
years to a limit of $V\sim24$ \citep{alc01}.
\end{itemize}
A comprehensive review of past and ongoing variability surveys can be found
in \citet{bec04}.

Recognizing the outstanding importance of variable objects, the last Decadal
Survey Report \citep{DSR01} highly recommended a major new initiative for
studying the variable sky, the Large Synoptic Survey Telescope (LSST;
\citealt{tys02,wal03}). The LSST\footnote{See \url[HREF]{http://www.lsst.org}}
will offer an unprecedented view of the faint variable sky: according to the
current designs it will scan the entire accessible sky every three nights to a
limit of $V\sim25$ with two observations per night in two different bands
(selected from a set of six). One of the LSST science goals\footnote{For more
details see \url[HREF]{http://www.lsst.org/Science/science\_goals.shtml}} will
be the exploration of the transient optical sky: the discovery and analysis of
rare and exotic objects (e.g. neutron star and black hole binaries), gamma-ray
bursts, X-ray flashes, and of new classes of transients, such as binary mergers
and stellar disruptions by black holes. The observed volume of space, and the
requirement to recognize and monitor these events --- in real time --- on a
``normally'' variable sky, will present a challenge to the project.

Since LSST will utilize\footnote{LSST will also use the $Y$ band at $\sim1$
$\mu m$. For more details see the LSST Science Requirement Document at
\url[HREF]{http://www.lsst.org/Science/lsst\_baseline.shtml}} the Sloan Digital
Sky Survey (SDSS; \citealt{yor00}) photometric system ($ugriz$,
\citealt{fuk96}), multiple photometric observations obtained by the SDSS
represent an excellent dataset for a pre-LSST study that characterizes the faint
variable sky and quantifies the variable population and its distribution in
magnitude-color-variability space. Here we present such a study of unresolved
sources in a region that has been imaged multiple times by the SDSS.

In Section~\ref{data} we give a brief overview of the SDSS imaging survey and
repeated scans of a $\sim290$ deg$^2$ region called Stripe 82. In
Section~\ref{analysis}, we describe methods used to select candidate variable
sources from the SDSS Stripe 82 data assembled, averaged and recalibrated by
\citet{ive07}, and present tests that show the robustness of the adopted
selection criteria. In the same section, we discuss the distribution of selected
variable sources in magnitude-color-variability space. The Milky Way halo
structure traced by selected candidate RR Lyrae stars is discussed in
Section~\ref{RRLyr}, and in Section~\ref{QSO} we estimate the fraction of
variable quasars. Implications for surveys such as the LSST are discussed in
Section~\ref{LSSTsec}, and our main results are summarized in
Section~\ref{discussion}.

\section{Overview of the SDSS Imaging and Stripe 82 Data\label{data}}

The quality of photometry and astrometry, as well as the large area covered by
the survey, make the SDSS stand out among available optical sky surveys
\citep{ses06}. The SDSS is providing homogeneous and deep ($r<22.5$) photometry
in five bandpasses ($u$, $g$, $r$, $i$, and $z$,
\citealt{gun98,hog02,smi02,gun06,tuc06}) accurate to 0.02 mag (root-mean-square
scatter, hereafter rms) for unresolved sources not limited by photon statistics
\citep{scr02,ive03a}, and with a zeropoint uncertainty of 0.02 mag
\citep{ive04a}. The survey sky coverage of 10,000 deg$^2$ in the northern
Galactic cap, and 300 deg$^2$ in the southern Galactic cap will result in
photometric measurements for well over 100 million stars and a similar number of
galaxies \citep{sto02}. The recent Data Release 5 \citep{amc07}\footnote{Please
see \url[HREF]{http://www.sdss.org/dr5}} lists photometric data for 215 million
unique objects observed in 8000 deg$^2$ of sky as part of the ``SDSS-I'' phase
that ran through June 2005. Astrometric positions are accurate to better than
$0.1\arcsec$ per coordinate (rms) for sources with $r<20.5$ \citep{pie03}, and
the morphological information from the images allows reliable star-galaxy
separation to $r\sim21.5$ \citep{lup02}. In addition, the 5-band SDSS photometry
can be used for very detailed source classification; e.g.~separation of quasars
and stars \citep{ric02}, spectral classification of stars to within 1-2 spectral
subtypes \citep{len98,fin00,haw02}, and even remarkably efficient color
selection of the horizontal branch and RR Lyrae stars \citep{yan00,sir04,ive05}
and low-metallicity G and K giants \citep{hel03}.

The equatorial Stripe 82 region (22h 24m~$< \alpha_{J2000} <$~04h 08m,
$-1.27\arcdeg<\delta_{J2000} < +1.27\arcdeg$, $\sim290$ deg$^2$), observed in
the southern Galactic cap, presents a valuable data source for variability
studies. The region was repeatedly observed (65 imaging runs by July 2005, but
not all cover the entire region), and it is the largest source of multi-epoch
data in the SDSS-I phase. Another source of the large number of scans is the
SDSS-II Supernova Survey \citep{fri07}. By averaging the repeated observations
of Stripe 82 sources, more accurate photometry than the nominal 0.02 mag
single-scan accuracy can be achieved. This motivated \citet{ive07} to produce a
catalog of recalibrated Stripe 82 observations. The catalog lists 58 million
photometric observations for 1.4 million unresolved sources that were observed
at least 4 times in each of the $gri$ bands (with a median of 10 observations
obtained over $\sim5$ years). The random photometric errors for PSF (point
spread function) magnitudes are below 0.01 mag for stars brighter than 19.5,
20.5, 20.5, 20, 18.5 in $ugriz$, respectively (about twice as accurate for
individual SDSS runs), and the spatial variation of photometric zeropoints is
not larger than $\sim$0.01 mag (rms). Following \citet{ive07}, we use PSF
magnitudes because they go deeper at a given signal-to-noise ratio than aperture
magnitudes, and have more accurate photometric error estimates than model
magnitudes. In addition, various low-order statistics such as root-mean-square
scatter ($\Sigma$), $\chi^2$ per degree of freedom ($\chi^2$), lightcurve
skewness ($\gamma$), minimum and maximum PSF magnitude, were computed for each
$ugriz$ band and each source. We compute $\chi^2$ per degree of freedom as
\begin{equation}
\chi^2 = \frac{1}{n-1}\sum_{i=1}^{n}\frac{(x_i-\langle x \rangle)^2}{\xi_i^2}
\end{equation}
and lightcurve skewness $\gamma$ as\footnote{We use equations from
\url[HREF]{http://www.xycoon.com/skewness\_small\_sample\_test\_1.htm}.}
\begin{equation}
\gamma = \frac{n^2}{(n-1)(n-2)}\frac{\mu_3}{\Sigma^3}
\end{equation}
\begin{equation}
\mu_3 = \frac{1}{n}\sum_{i=1}^{n}(x_i-\langle x \rangle)^3
\end{equation}
\begin{equation}
\Sigma = \sqrt{\frac{1}{n-1}\sum_{i=1}^{n}(x_i-\langle x \rangle)^2}
\end{equation}
where $n$ is the number of detections, $x_i$ is the magnitude,
$\langle x \rangle$ is the mean magnitude, and $\xi_i$ is the photometric error.

Separation of quasars and stars, as well as efficient color selection of
horizontal branch and RR Lyrae stars, depend on accurate $u$ band photometry. To
ensure this, we select 748,084 unresolved sources from the \citet{ive07} catalog
with at least 4 detections in the $u$ band. A catalog of variable sources
selected from this sample is analyzed in Section~\ref{analysis} below.

\section{Analysis of Stripe 82 Catalog of Variable Sources\label{analysis}}

In this section we describe methods for selecting candidate variable sources,
and present tests that show the robustness of the adopted selection criteria.
The distribution of selected variable sources in magnitude-color-variability
space is also presented.

\subsection{Methods and Selection Criteria\label{methods}}

Due to a relatively small number of observations per source and random sampling,
we do not perform lightcurve fitting, but instead use low order statistics to
select candidate variables and study their properties. There are four parameters
(median PSF magnitude, root-mean-square scatter $\Sigma$, $\chi^2$, and
lightcurve skewness $\gamma$) measured in five photometric bands ($u$, $g$, $r$,
$i$, and $z$), for a total of 20 parameters. In the analysis presented here, we
utilize eight of them:
\begin{itemize}
\item
median PSF magnitudes in the $ugr$ bands (corrected for interstellar extinction
using the map from \citealt{SFD98}) because the $g-r$ vs.~$u-g$ color-color
diagram has the most classification power (e.g.~\citealt{smo04} and references
therein).
\item
$\Sigma$ and $\chi^2$ in the $g$ and $r$ bands, and
\item
lightcurve skewness $\gamma(g)$ (the $g$ band combines a high signal-to-noise
ratio and large variability amplitude for the majority of variable sources).
\end{itemize}

\begin{figure}
\epsscale{1}
\plotone{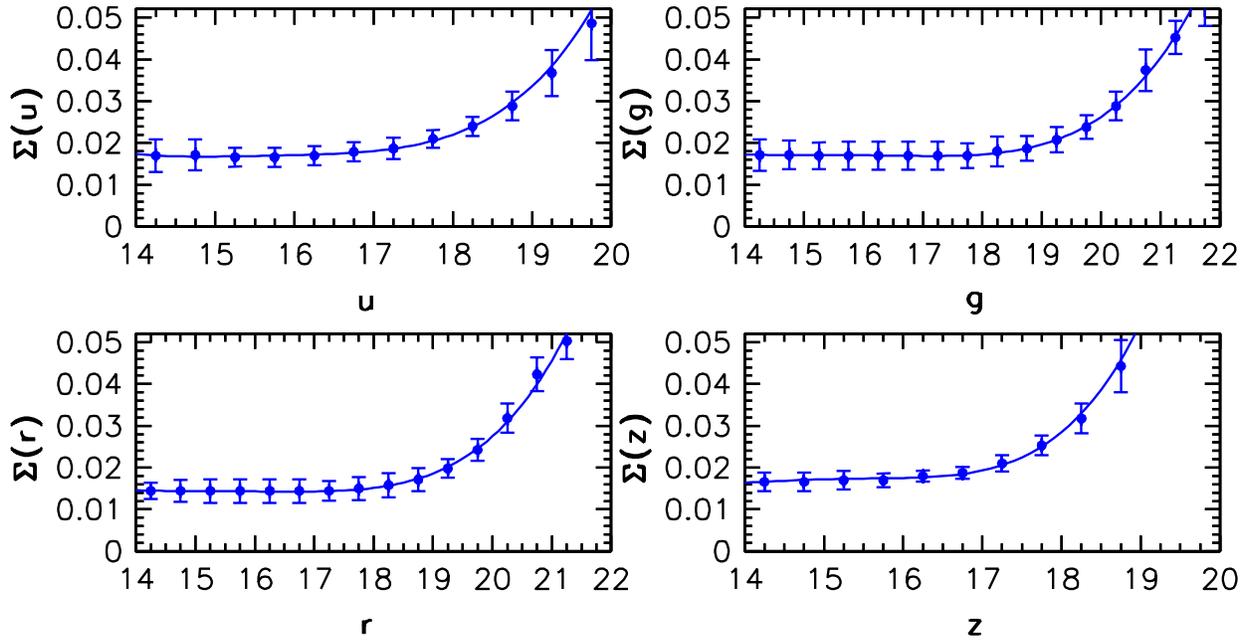}
\caption{
The dependence of the median root-mean-square (rms) scatter $\Sigma$ in SDSS
$ugrz$ bands on magnitude (symbols). The vertical bars show the rms scatter of
$\Sigma$ in each bin (not the error of the median). The dependence of $\Sigma$
in the $i$ band is similar to the $r$ band dependence. In each band, a
fourth-degree polynomial is fitted through medians and shown by the solid line.
\label{phot_acc}}
\end{figure}

The observed root-mean-square scatter $\Sigma$ includes both the intrinsic
variability $\sigma$ and the mean photometric error $\langle \xi(m) \rangle$ as
a function of magnitude. The dependence of $\Sigma$ on magnitude in the $ugriz$
bands, is shown in Figure~\ref{phot_acc}. For sources brighter than 18, 19.5,
19.5, 19, and 17.5 mag in the $ugriz$, respectively, the SDSS delivers $2\%$
photometry with little or no dependence on magnitude. We determine
$\langle \xi(m) \rangle$ by fitting a fourth degree polynomial to median
$\Sigma$ values in 0.5 mag wide bins (here we assume that the majority of
sources are not variable). The theoretically expected $\langle \xi(m) \rangle$
function \citep{str01}
\begin{equation}
\langle \xi(m) \rangle = a+b10^{0.4m}+c10^{0.8m}
\end{equation}
provides equally good fits. We define the intrinsic variability $\sigma$
(hereafter rms scatter $\sigma$) as
\begin{equation}
\sigma = (\Sigma^2-\langle \xi(m) \rangle^2)^{1/2}
\label{sigma}
\end{equation}
for $\Sigma>\langle \xi(m) \rangle$, and $\sigma=0$ otherwise.

As the first variability selection criterion, we adopt
$\sigma(g) \geqslant 0.05$ mag and $\sigma(r) \geqslant 0.05$ mag (hereafter
written as $\sigma(g,r) \geqslant 0.05$ mag). At the bright end, this criterion
is equivalent to selecting sources with rms scatter greater than $2.5\sigma_0$,
where $\sigma_0 = 0.02$ mag is the measurement noise. Selection cuts are applied
simultaneously in the $g$ and $r$ bands to reduce the number of ``false
positives'' (intrinsically non-variable sources selected as candidate variable
sources due to measurement noise). About $6\%$ of sources pass the $\sigma$ cut
in each band separately, and $\sim3\%$ of sources pass the cut in both bands
simultaneously. By selecting sources with $\sigma(g,r) \geqslant 0.05$ mag, we
also select faint sources that have large $\sigma$ due to large photometric
errors at the faint end. To only select faint sources with statistically
significant rms scatter, we apply the $\chi^2$ test as the second selection cut.

\begin{figure}
\epsscale{1}
\plotone{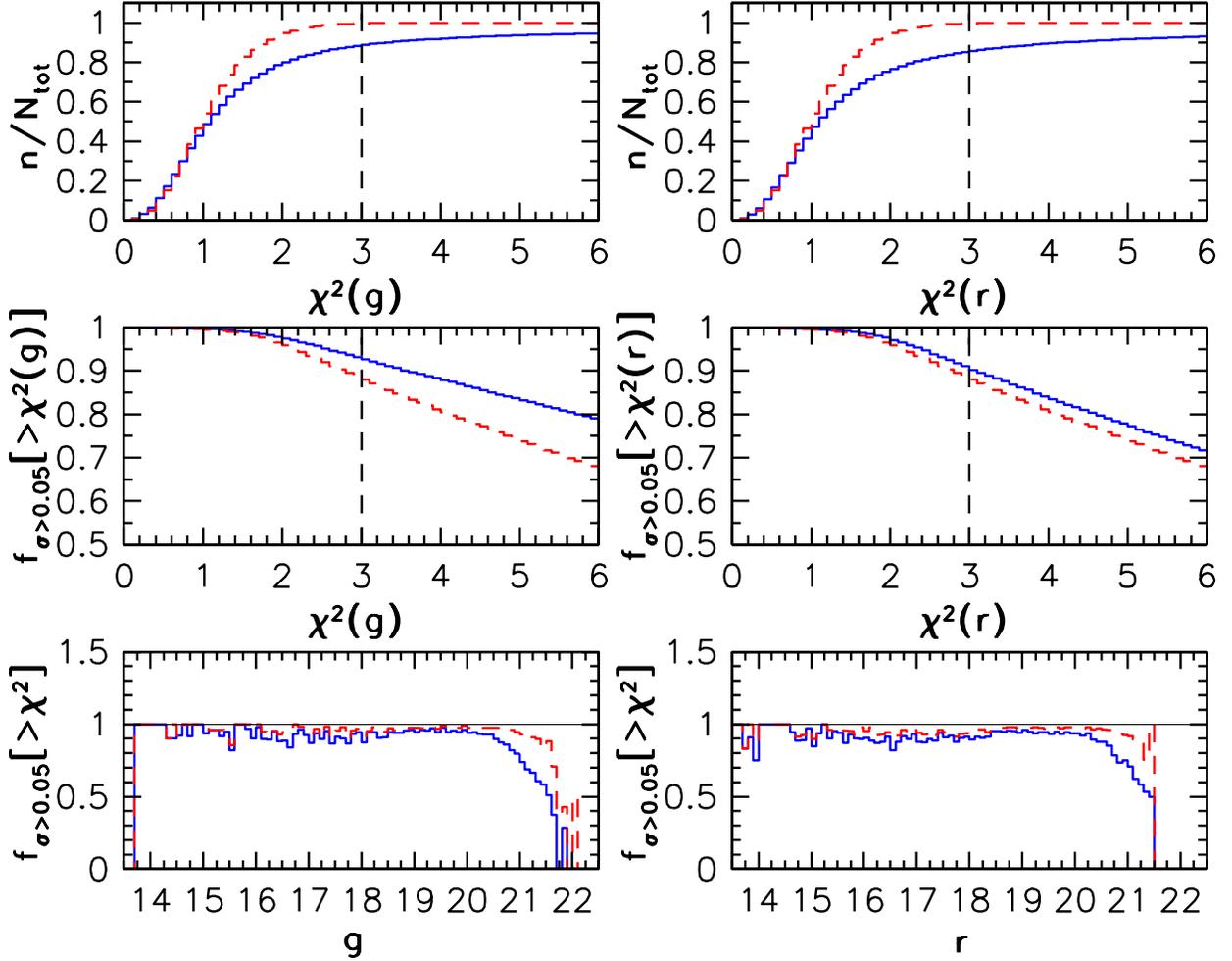}
\caption{
{\em Top}: The cumulative distribution of $\chi^2$ $g$ and $r$ values for
all sources (solid line) and a reference Gaussian $\chi^2$ distribution with
9 degrees of freedom (dashed line). Vertical dashed lines show adopted selection
cuts on $\chi^2(g)$ and $\chi^2(r)$ values. {\em Middle}: The fraction of
$\sigma(g,r)\geqslant0.05$ mag sources with $\chi^2$ per degree of freedom
greater than $\chi^2$ (only in the $g$ or $r$ band: solid line, both in the $g$
and $r$ bands: dashed line). {\em Bottom}: The fraction of
$\sigma(g,r)\geqslant0.05$ mag sources with $\chi^2(m)\geqslant2$ (dashed line)
or $\chi^2(m)\geqslant3$ (solid line) as a function of magnitude for
$m=g,r$ bands, respectively.
\label{chi_cum}}
\end{figure}

In the $\chi^2$ test, the value of $\chi^2$ per degree of freedom (calculated
with respect to a weighted mean magnitude and using errors computed by the
photometric pipeline) determines whether the observed lightcurve is consistent
with the Gaussian distribution of errors. Large $\chi^2$ values show that the
rms scatter is inconsistent with random fluctuations. \citet{ive03a,ive07}
used multi-epoch SDSS observations to show that the photometric error
distribution in the SDSS roughly follows a Gaussian distribution. A comparison
of $\chi^2$ distributions in the $g$ and $r$ bands with a reference Gaussian
$\chi^2$ distribution is shown in Figure~\ref{chi_cum}. As evident, $\chi^2$
distributions in both bands roughly follow the reference Gaussian $\chi^2$
distribution for $\chi^2<1$, demonstrating that median photometric errors are
correctly determined. The discrepancy for larger $\chi^2$ is due to variable
sources rather than non-Gaussian error distributions, as we demonstrate below.

The second selection cut, $\chi^2(g) \geqslant 3$ and $\chi^2(r) \geqslant 3$
(hereafter written as $\chi^2(g,r) \geqslant 3$), selects $\sim90\%$ of
$\sigma(g,r) \geqslant 0.05$ mag sources, as shown in Figure~\ref{chi_cum}
(middle panels). The effectiveness of the $\chi^2$ test is demonstrated in the
bottom panel of Figure~\ref{chi_cum}. For magnitudes fainter than $g=20.5$, the
fraction of candidate variables decreases as photometric errors increase. The
selection is relatively uniform for sources brighter than $g=20.5$, and we adopt
this value as the flux limit for the selected variable sample.

There are 662,195 sources brighter than $g=20.5$ in the full sample. Using
$\sigma(g,r) \geqslant 0.05$ mag and $\chi^2(g,r) \geqslant 3$ as the selection
criteria, we select 13,051 candidate variable sources\footnote{A list of
candidate variable sources and their data from \citet{ive07} are publicly
available from
\url[HREF]{http://www.sdss.org/dr5/products/value\_added/index.html}}.
Therefore, {\em at least $2\%$ of unresolved optical sources brighter than
$g=20.5$ appear variable at the $\geqslant0.05$ mag level (rms) simultaneously
in the $g$ and $r$ bands}. The fraction of selected variable sources is not a
strong function of the minimum required number of observations, but it does
depend on the stellar density because the number of stars increases at lower
Galactic latitudes (see Fig.~5 in \citealt{ive07}) while the quasar count
remains the same.

\subsection{The Counts of Variable Sources\label{counts}}

In this section we estimate the completeness and efficiency of the candidate
variable sample, and discuss the dependence of counts, rms scatter,
$\sigma(g)/\sigma(r)$ ratio, and the lightcurve skewness $\gamma(g)$ on the
position in the $g-r$ vs.~$u-g$ color-color diagram.

\subsubsection{Completeness}

The selection completeness, defined as the fraction of true variable sources
recovered by the algorithm, depends on the lightcurve shape and amplitudes. Due
to a fairly large number of observations (median of 10), and small
$\sigma(g,r)$ cutoff compared to typical amplitudes of variable sources
(e.g. most RR Lyrae stars and quasars have peak-to-peak amplitudes $\sim1$ mag),
we expect the completeness to be fairly high for RR Lyrae stars ($\ga95\%$, see
Section~\ref{RRLyr}) and quasars ($\sim90\%$, see Section~\ref{QSO}). The
completeness for other types of variable sources, such as flares and eclipsing
binaries, is hard to estimate, but is probably low due to sparse sampling.

\subsubsection{Efficiency\label{efficiency}}

The selection efficiency, defined as the fraction of true variable sources
in the candidate variable sample, determines the robustness of the selection
algorithm. The main diagnostic for the robustness of the adopted selection
criteria is the distribution of selected candidates in the SDSS color-magnitude
and color-color diagrams. The position of a source in these diagrams is a good
proxy for its spectral classification \citep{len98,fan99,fin00,smo04}.

\begin{figure}
\epsscale{0.35}
\plotone{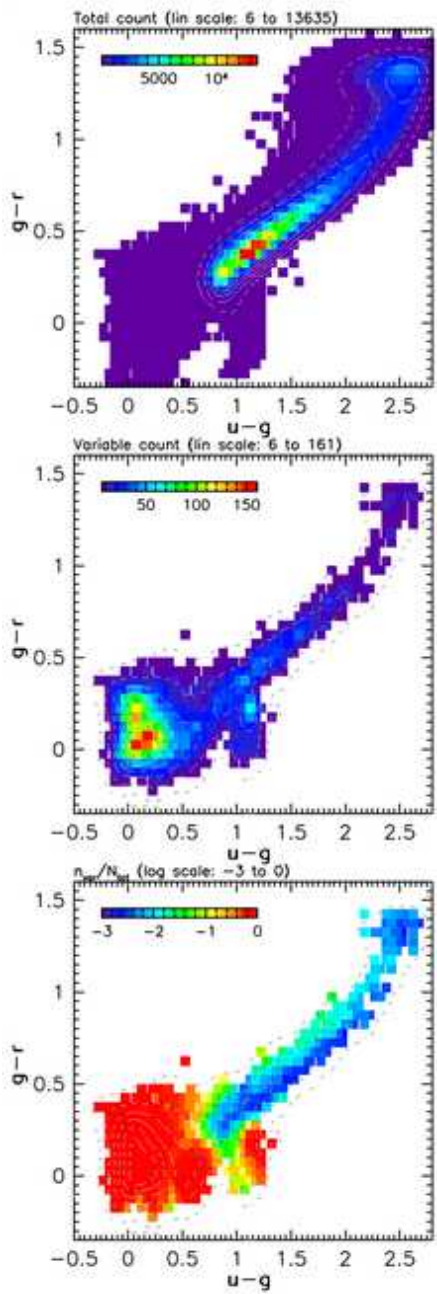}
\caption{
The distribution of counts for the full sample (top), candidate variable sample
(middle), and the ratio of two counts (bottom) in the $g-r$ vs.~$u-g$
color-color diagram for sources brighter than $g=20.5$, binned in 0.05 mag
bins. Contours outline distributions of unbinned counts. Note the remarkable
difference between the distribution of all sources and that of the variable
sample, which demonstrates that the latter are robustly selected.
\label{map_counts}}
\end{figure}

Figure~\ref{map_counts} compares the distribution of candidate variable sources
to that of all sources in the $g-r$ vs.~$u-g$ color-color diagram. Were the
selection a random process, the selected candidates would have the same
distribution as the full sample. The distributions of candidate variables and
of the full sample are remarkably different, demonstrating that the candidate
variables are {\it not} randomly selected from the parent sample. 

The three dominant classes of variable objects are quasars, RR Lyrae stars, and
stars from the main stellar locus. The most obvious difference between the
variable and the full sample distributions is a much higher fraction of
low-redshift quasars ($<2.2$, recognized by their UV excess, $u-g < 0.7$, see
\citealt{ric02}) and RR Lyrae stars ($u-g \sim 1.15$, $g-r < 0.3$, see
\citealt{ive05}) in the variable sample, and vividly shown in the bottom panel
of Figure~\ref{map_counts}.

Another interesting feature visible in this panel is a gradient in the fraction
of variable main stellar locus stars (perpendicular to the main stellar locus).
We investigate this gradient by first defining principal colors
\begin{equation}
P_1 = 0.91u-0.495g-0.415r-1.28
\end{equation}
and
\begin{equation}
s = -0.249u+0.794g-0.555r+0.234
\end{equation}
where $P_1$ and $s$ are principal axis parallel and perpendicular to the main
stellar locus, respectively \citep{ive04a}. The $s$ color is a measure of
metallicity \citep{len98}, and $s>0.05$ stars are expected to be metal poor
\citep{hel03}. Sources with $r<19$ and $0<P_1<0.9$ are selected and binned in
four $s$ bins. For each bin we calculate the fraction of source with
$\sigma(g)\geqslant0.05$ mag, the fraction of variable sources (selected with
$\sigma(g,r) \geqslant 0.05$ mag and $\chi^2(g,r) \geqslant 3$), median
$\sigma(g)$, and the total number of sources in the bin (see
Table~\ref{principal}). A greater fraction of variable sources in the last bin
($s>0.06$) indicates that, on average, metal-poor main stellar locus stars
are more variable than the metal-rich stars. This could be because this sample
of metal-poor stars is expected to have a high fraction of giants.

\begin{figure}
\epsscale{0.8}
\plotone{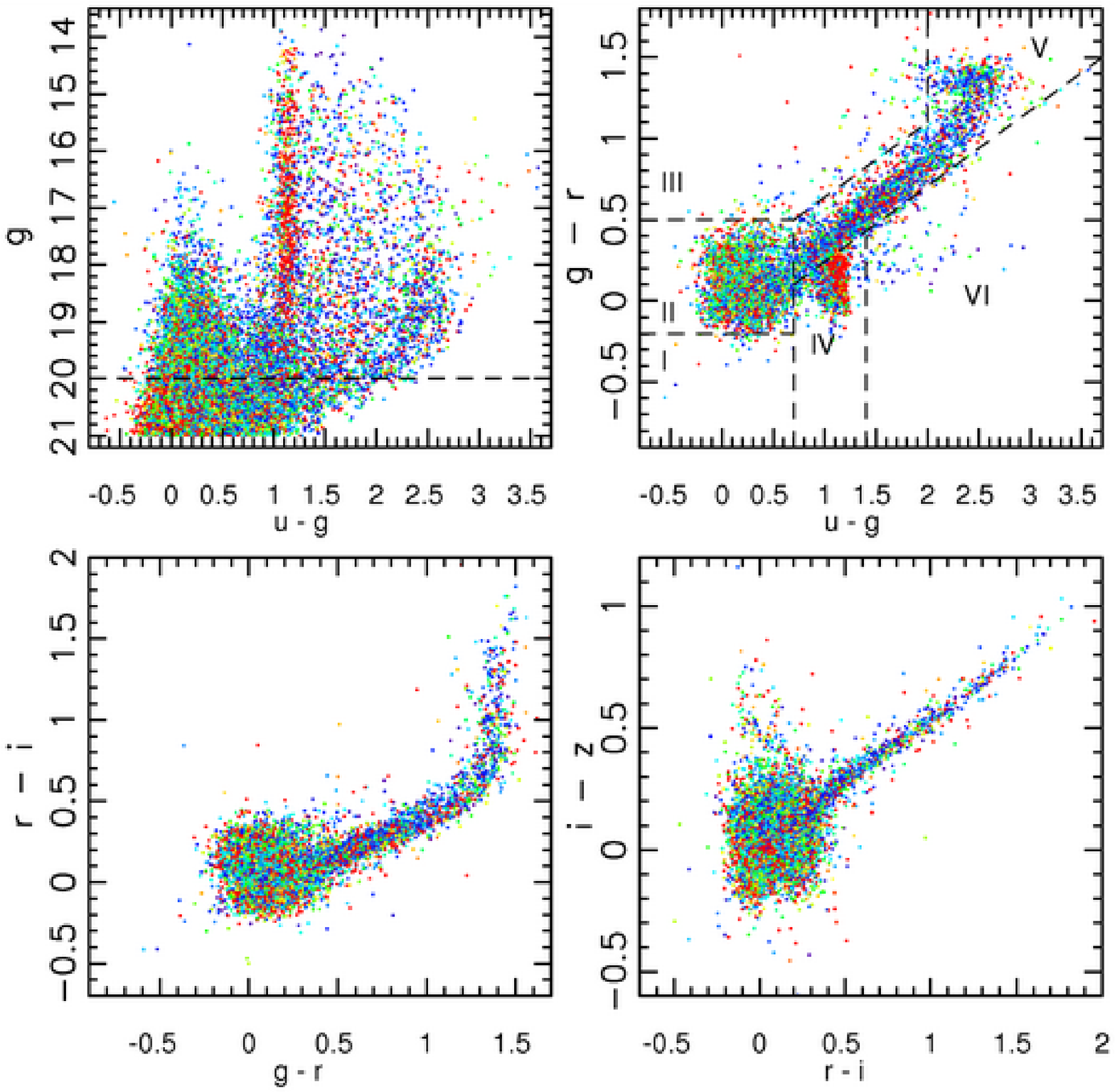}

\epsscale{0.3}
\plotone{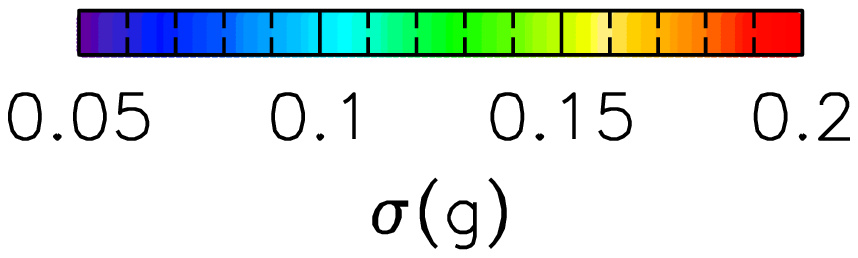}
\caption{
The distribution of 18,329 candidate variable sources brighter than $g=21$ in
representative SDSS color-magnitude and color-color diagrams. Candidate
variables are color-coded by their rms scatter in the $g$ band
(0.05-0.2, see the legend, red if larger or equal than 0.2). Only sources
brighter than $g=20$ are plotted in color-color diagrams. Note how RR Lyrae
stars ($u-g\sim1.15$, red dots, $\sigma(g)\ga0.2$ mag) and low-redshift quasars
($u-g\leqslant0.7$, green dots, $\sigma(g)\ga0.1$ mag) stand out as highly
variable sources. The regions marked in the top right panel are used for
quantitative comparison of the overall and variable source distributions (see
Table \ref{regions}).
\label{rmsg_col}}
\end{figure}

In order to quantify the differences between the full and the variable sample,
we follow \citet{ses06} and divide the $g-r$ vs.~$u-g$ color-color diagram into
six characteristic regions, each dominated by a particular type of source, as
shown in Figure~\ref{rmsg_col}. The fractions and counts of variable and all
sources in each region are listed in Table~\ref{regions} for $g<19$, $g<20.5$,
and $g<22$ flux-limited samples. Notably, in the adopted $g<20.5$ flux limit,
the fraction of Region II sources (dominated by low-redshift quasars) in the
variable sample is $63\%$, or $\sim30$ times greater than the fraction of
Region II sources in the full sample ($\sim2\%$). The fraction of Region IV
sources (which include RR Lyrae stars) in the variable sample is also high when
compared to the full sample ($\sim6$ times higher).

As shown in Table~\ref{regions}, in the $g=20.5$ flux-limited sample, we
find that low-redshift quasars and RR Lyrae stars (i.e. Regions II and IV)
make $70\%$ of the variable population, while representing only $3\%$ of all
sources. Quasars alone account for $63\%$ of the variable population. Stars from
the main stellar locus represent $95\%$ of all sources and $25\%$ of the
variable sample: about $0.5\%$ of the stars from the locus are variable at the
$\geqslant0.05$ mag level.

\subsection{The Properties of Variable Sources\label{properties}}

Various lightcurve properties, such as shape and amplitude, are expected to be
correlated with stellar types. In this section we study the distribution of the
rms scatter in the $u$ and $g$ bands, and $\sigma(g)/\sigma(r)$ ratio as a
function of the $u-g$ and $g-r$ colors. To emphasize trends, we bin sources and
present median values for each bin.

\begin{figure}
\epsscale{0.85}
\plotone{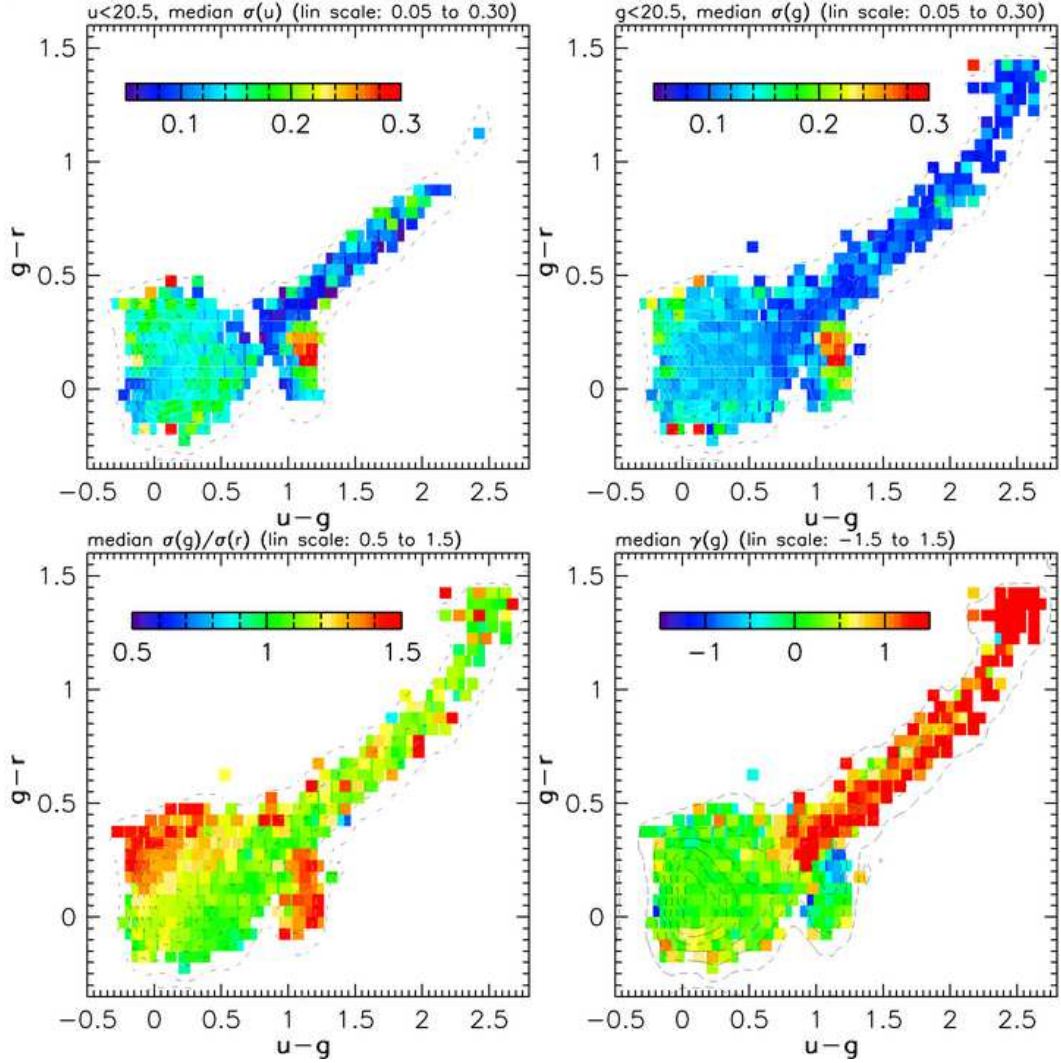}
\caption{
The distribution of the rms scatter $\sigma(u)$ (top left), rms scatter
$\sigma(g)$ (top right), $\sigma(g)/\sigma(r)$ ratio (bottom left), and
$\gamma(g)$ (bottom right) for the variable sample in the $g-r$ vs.~$u-g$
color-color diagram. Sources are binned in 0.05 mag wide bins and the median
values are color-coded. Color ranges are given at the top of each panel, going
from blue to red, where the green color is in the mid-range. Values outside the
range saturate in blue or red. Contours outline the count distributions on a
linear scale in steps of $15\%$. Flux limit is $g<20.5$, with an additional
$u<20.5$ limit in the top left panel. {\em Bottom left:} On average, RR Lyrae
stars have $\sigma(g)/\sigma(r)\sim1.4$, main stellar locus stars have
$\sigma(g)/\sigma(r)\sim1$, while low-redshift quasars show a gradient of
$\sigma(g)/\sigma(r)$ values. {\em Bottom right:} On average, quasars and $c$
type RR Lyrae stars ($u-g\sim1.15$, $g-r<0.15$) have $\gamma(g)\sim0$, $ab$ type
RR Lyrae ($u-g\sim1.15$, $g-r>0.15$) have negative skewness, and stars in the
main stellar locus have positive skewness.
\label{med_rms}}
\end{figure}

The distribution of the median $\sigma(u)$ and $\sigma(g)$ values in the $g-r$
vs.~$u-g$ color-color diagram is shown in the top two panels of
Figure~\ref{med_rms}. RR Lyrae stars show larger rms scatter ($\ga 0.3$ mag) in
the $u$ and $g$ bands, than low-redshift quasars or stars from the main stellar
locus. Quasars also show slightly larger rms scatter in the $u$ band
($\sim0.1$ mag) than in the $g$ band ($\sim0.07$ mag), as discussed by
\citet{kin91},\citet{ive04b}, and \citet{VB04}. If we define the degree of
variability as the root-mean-square scatter in the $g$ band, then on average RR
Lyrae stars show the greatest variability, followed by quasars and the main
stellar locus stars.

Another distinctive characteristic of variable sources is the ratio of flux
changes in different bandpasses. This property can be used to select different
types of variable sources. For example, RR Lyrae stars are bluer when brighter,
a behavior used by \citet{ive00} to select RR Lyrae using 2-epoch SDSS data.
Here we define a new parameter, $\sigma(g)/\sigma(r)$, to express the ratio of
flux changes in the $g$ and $r$ bands, and study its distribution in the $g-r$
vs.~$u-g$ color-color diagram. In particular, we examine this distribution and
its median values for three dominant classes of variable sources: quasars, RR
Lyrae stars, and stars from the main stellar locus.

The bottom left panel in Figure~\ref{med_rms} shows the distribution of median
$\sigma(g)/\sigma(r)$ values as a function of $u-g$ and $g-r$ colors.
Using Fig.~\ref{med_rms} we note that on average:
\begin{itemize}
\item
RR Lyrae stars have $\sigma(g)/\sigma(r)\sim1.4$
\item
Main stellar locus stars have $\sigma(g)/\sigma(r)\sim1$, and
\item
Quasars show a $\sigma(g)/\sigma(r)$ gradient in the $g-r$ vs.~$u-g$ color-color
diagram.
\end{itemize}

\begin{figure}
\epsscale{0.9}
\plotone{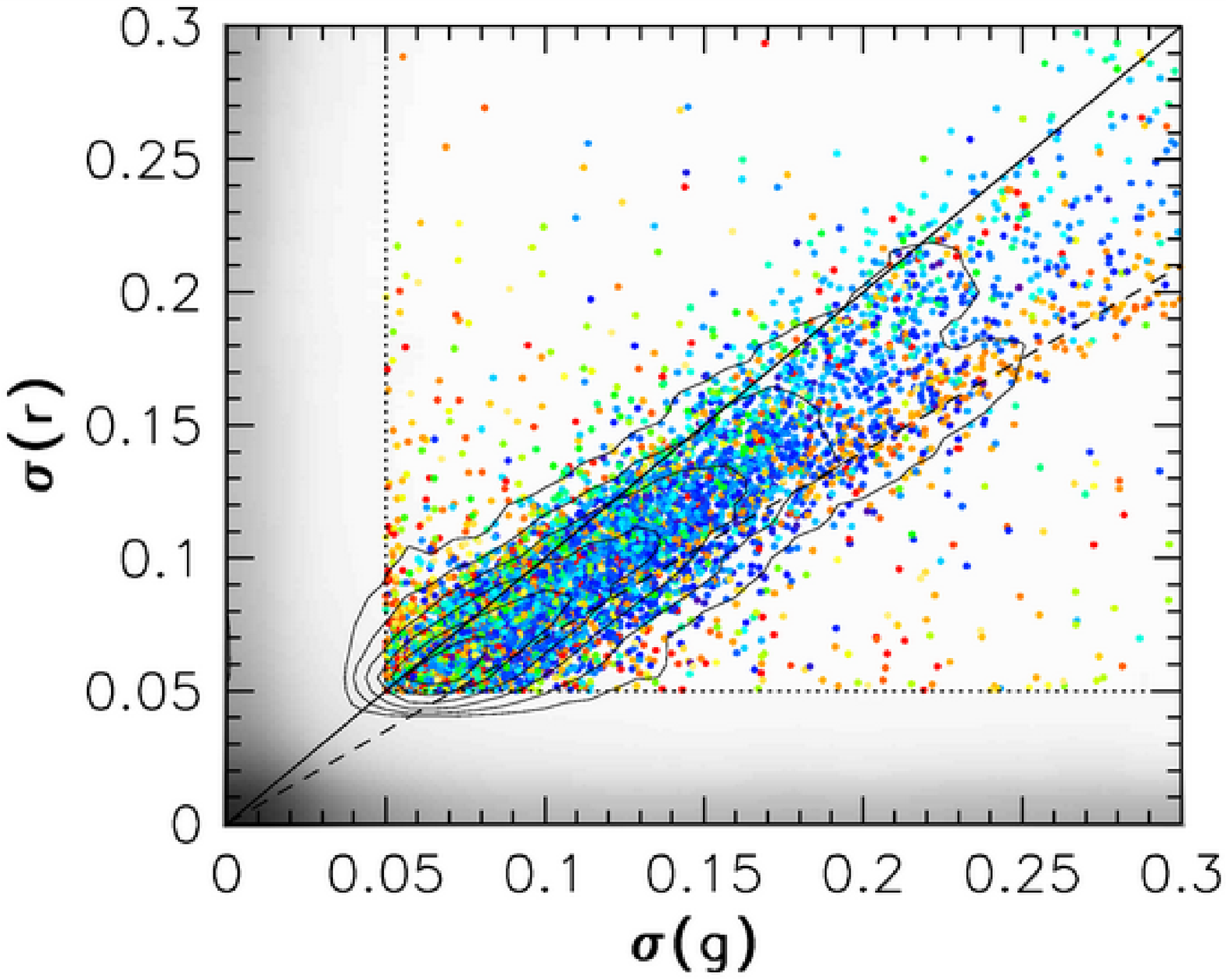}

\plottwo{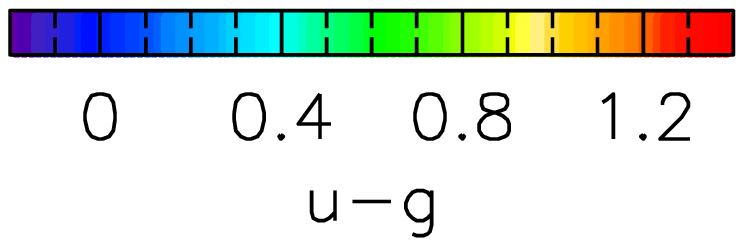}{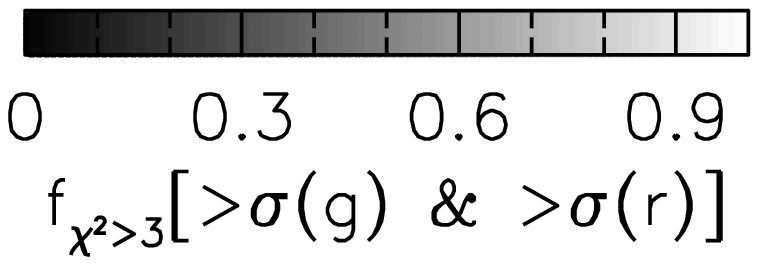}
\caption{
The distribution of candidate variable sources in the $g<20.5$ flux-limited
sample is shown by linearly-spaced contours, and by symbols color-coded by the
$u-g$ color for sources with $\sigma(g)\geqslant0.05$ mag and
$\sigma(r)\geqslant0.05$ mag. The dotted lines show the adopted $\sigma(g,r)$
selection cut. The thick solid line shows $\sigma(g)=\sigma(r)$, while the
dashed line shows $\sigma(g)=1.4\sigma(r)$ relation representative of RR Lyrae
stars. Note that sources following the $\sigma(g)=1.4\sigma(r)$ relation tend to
have $u-g\sim1$, as expected for RR Lyrae stars. The greyscale background shows
the fraction of $\chi^2(g,r)\geqslant 3$ sources which also have
$\sigma(g)\geqslant x$ and $\sigma(r)\geqslant y$ and demonstrates that large
$\chi^2$ sources also have large $\sigma$.
\label{plot_Bayes}}
\end{figure}

The average value of $\sigma(g)/\sigma(r)\sim1.4$ in Region IV indicates that RR
Lyrae stars dominate the variable source count in this region. The ratio of 1.4
for RR Lyrae stars was also previously found by \citet{ive00}. While
Figure~\ref{med_rms} only presents median values of the rms scatter,
Figure~\ref{plot_Bayes} shows how the rms scatter in the $g$ and $r$ bands
correlates with the $u-g$ color for individual sources. Variable sources that
follow the $\sigma(g)=1.4\sigma(r)$ relation also correlate with the $u-g$
color, and have $u-g\sim1$, as expected for RR Lyrae stars.

The average ratio of $\sigma(g)/\sigma(r)\sim1$ (i.e. gray flux variations) for
stars in the main stellar locus suggests that the variability could be caused by
eclipsing systems. The distribution of $\gamma(g)$ for main stellar locus stars
further strengthens this possibility, as discussed in Section~\ref{skewness}
below.

\begin{figure}
\epsscale{0.7}
\plotone{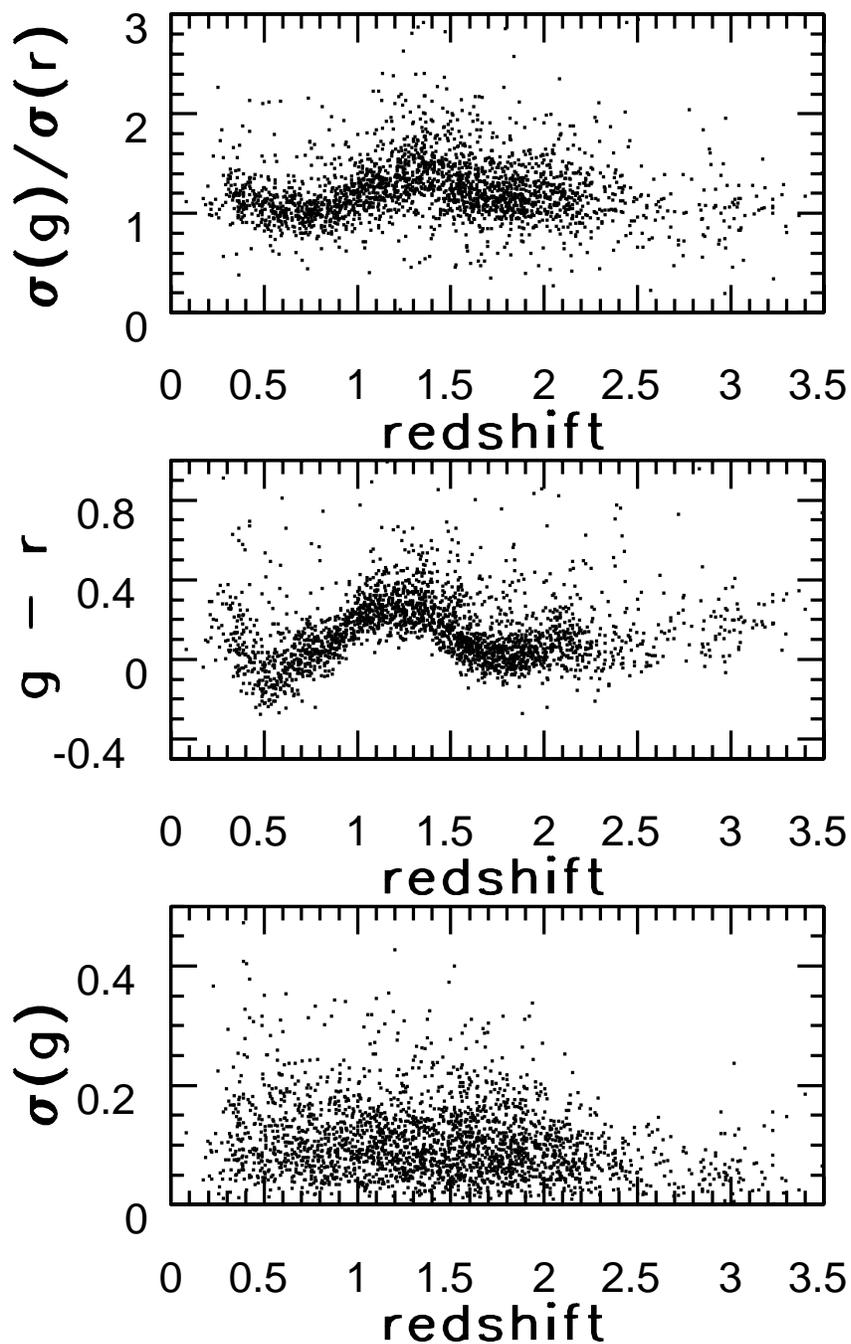}
\caption{
The dependence of $\sigma(g)/\sigma(r)$ (top), $g-r$ (middle), and $\sigma(g)$
on redshift for a sample of spectroscopically confirmed unresolved quasars from
\citet{sch05}. The $\sigma(g)/\sigma(r)$ gradient shown in Fig.~\ref{med_rms}
(bottom left panel) can be explained by the local maximum of
$\sigma(g)/\sigma(r)$ in the 1.0 to 1.6 redshift range.
\label{qso_redshift}}
\end{figure}

The gradient in the $\sigma(g)/\sigma(r)$ ratio observed for low-redshift
quasars in the $g-r$ vs.~$u-g$ color-color diagram suggests that the variability
correlation between the $g$ and $r$ bands is more complex than in the case of RR
Lyrae or main stellar locus stars. \citet{wil06} show that the photometric
color changes for quasars depend on the combined effects of continuum changes,
emission-line changes, redshift, and the selection of photometric bandpasses.
They note that due to the lack of variability of the lines, measured photometric
color is not always bluer in brighter phases, but depends on redshift and the
filters used. To verify the dependence of broad-band photometric variability on
redshift, we plot $\sigma(g)/\sigma(r)$ vs.~redshift for all spectroscopically
confirmed unresolved quasars from \citet{sch05} which are in Stripe 82, as shown
in Figure~\ref{qso_redshift}. We confirm that the broad-band photometric
variability depends on the redshift, and that the $\sigma(g)/\sigma(r)$ gradient
in the $g-r$ vs.~$u-g$ color-color diagram can be explained by the increase in
$\sigma(g)/\sigma(r)$ from $\sim1$ to $\sim1.6$ in the 1.0 to 1.6 redshift
range. This effect is due to the Mg II emission line (more stable in flux than
the continuum) moving through the $r$ band filter over this redshift range. The
implied correlation of the $u-g$ and $g-r$ colors with redshift is consistent
with the discussion by \citet{ric02}. The lack of noticeable correlation of
$\sigma(g)$ with redshift is due to the combined effects of the dependence of
$\sigma(g)$ on the rest-frame wavelength and time which cancel out (for a
detailed model see \citealt{ive04b}).

\subsection{Skewness as a Proxy for Dominant Variability Mechanism\label{skewness}}

Lightcurve skewness, a measure of the lightcurve asymmetry, provides additional
information on the type of variability. Negatively skewed, asymmetric
lightcurves indicate variable sources that spend more time fainter than
$(m_{min}+m_{max})/2$, where $m_{min}$ and $m_{max}$ are magnitudes at the
minimum and maximum. Type $ab$ RR Lyrae stars, for example, have negatively
skewed lightcurves ($\gamma\sim-0.5$, \citealt{wlb06}). Positively skewed,
asymmetric lightcurves indicate variable sources that spend more time brighter
than $(m_{min}+m_{max})/2$ (e.g. eclipsing systems). Sources with symmetric
lightcurves will have $\gamma\sim0$.

The bottom right panel in Figure~\ref{med_rms} shows the distribution of the
median $\gamma(g)$ as a function of the position in the $g-r$ vs.~$u-g$
color-color diagram. On average, quasars and $c$ type RR Lyrae stars
($u-g\sim1.15$, $g-r<0.15$) have $\gamma(g)\sim0$, $ab$ type RR Lyrae
($u-g\sim1.15$, $g-r>0.15$) have negative skewness ($\gamma(g)\sim-0.5$), and
stars in the main stellar locus have positive skewness.

\begin{figure}
\epsscale{0.3}
\plotone{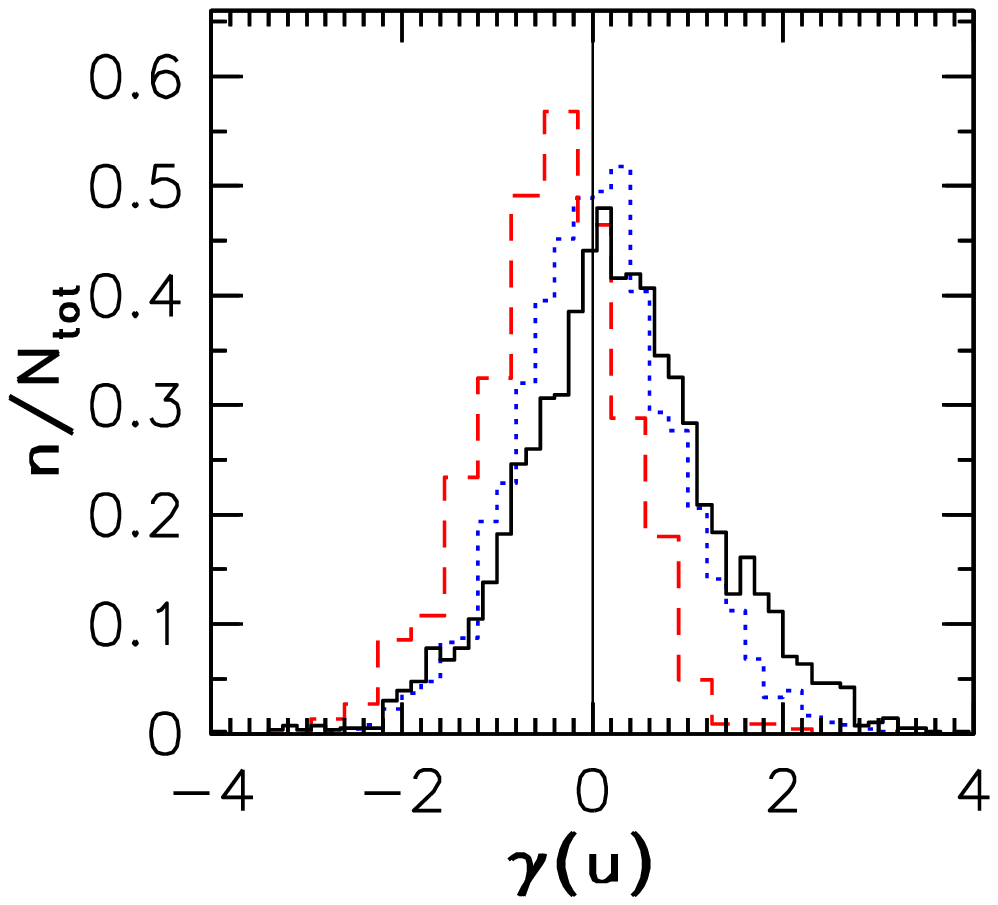}

\plotone{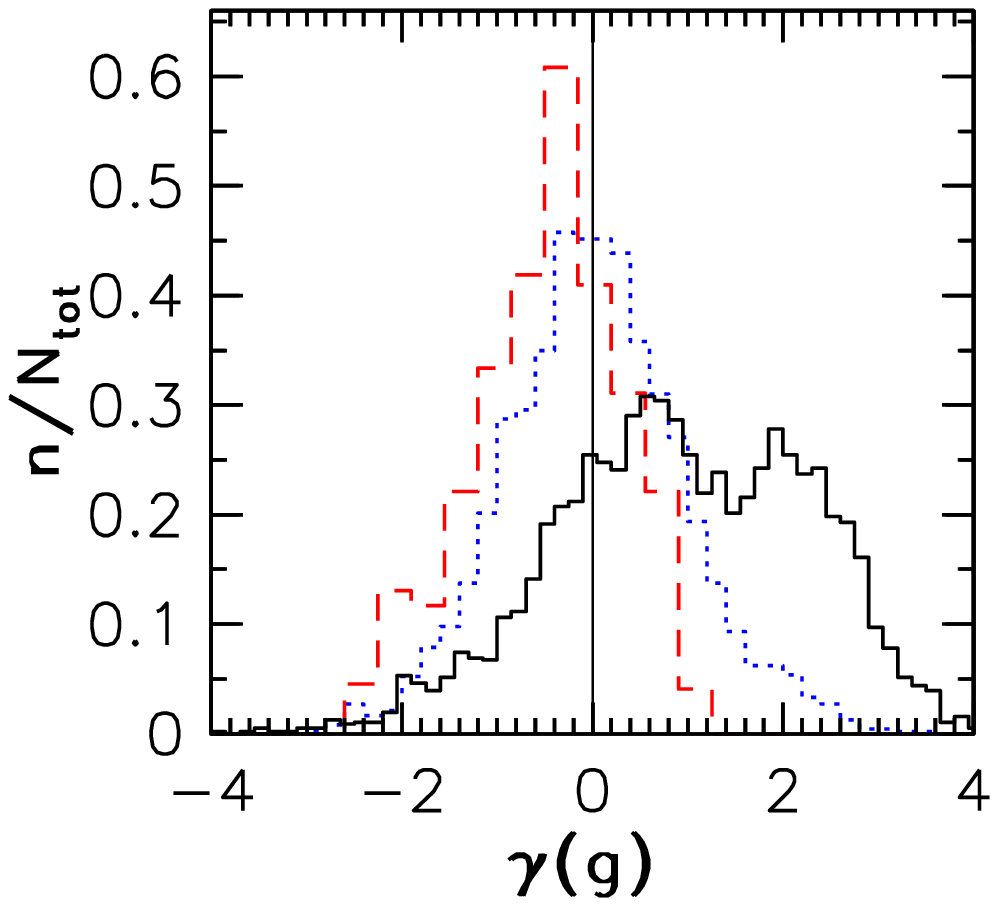}

\plotone{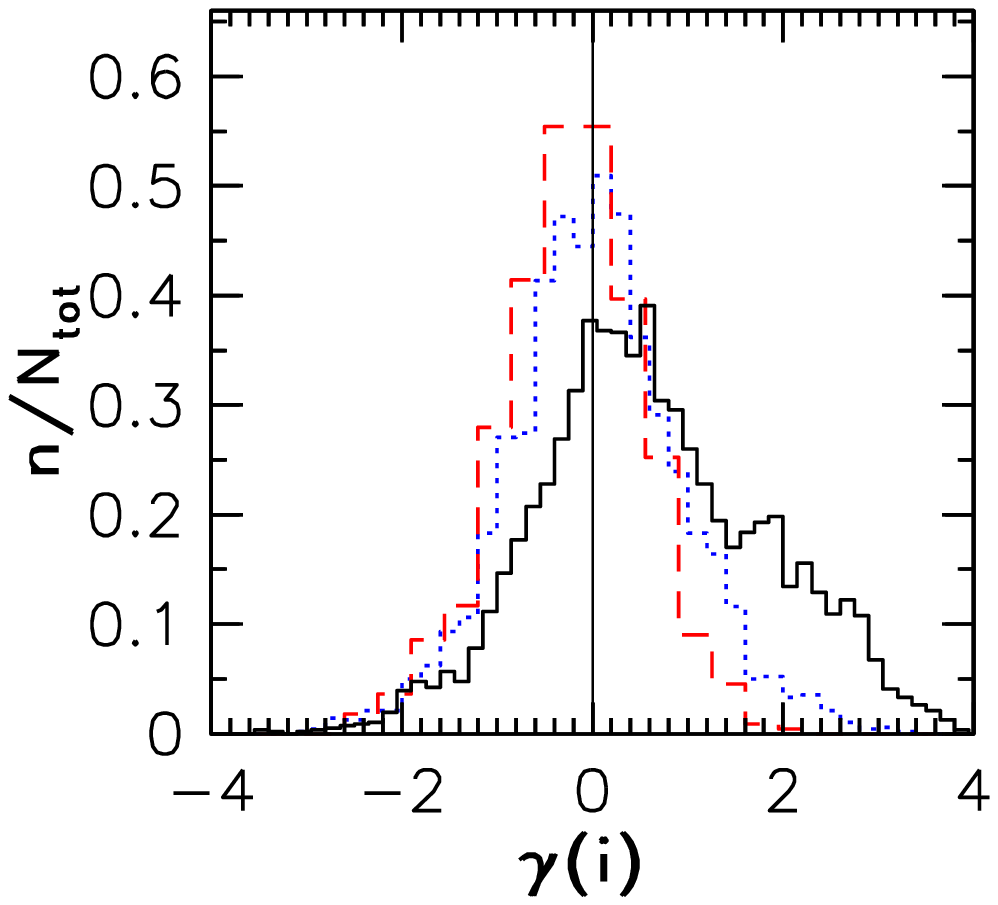}
\caption{
The lightcurve skewness distribution in the $ugi$ bands for spectroscopically
confirmed unresolved quasars (dotted line), candidate RR Lyrae stars (dashed
line), and variable main stellar locus stars (solid line, Region V, see
Fig.~\ref{rmsg_col} for the definition). The distribution of the skewness in the
$r$ band is similar to the $g$ band distribution, and the distribution of
skewness in the $z$ band is similar to the $u$ band distribution (therefore the
$r$ and $z$ data are not shown). Stars in the main stellar locus show bimodality
in the $\gamma(g)$ suggesting at least two, and perhaps more, different
populations of variables. Similar bimodality is also discernible in the $r$
band, while it is less pronounced in the $i$ band and not detected in the $u$
and $z$ bands. Quasars have symmetric lightcurves ($\gamma\sim0$) and their
distribution of skewness does not change between bands. 
\label{skew_hist}}
\end{figure}

Figure~\ref{skew_hist} shows the distribution of the lightcurve skewness in the
$ugi$ bands for spectroscopically confirmed unresolved quasars from
\citet{sch05} which are in Stripe 82, candidate RR Lyrae stars (selection
details are discussed in Section~\ref{RRLyr} below), and main stellar locus
stars from our variable sample. Stars in the main stellar locus show a bimodal
$\gamma(g)$ distribution. This distribution suggests at least two, and perhaps
more, different populations of variables. Indeed, when spectroscopically
confirmed M dwarfs are selected, a third peak appears at $\gamma(g)~-2.5$,
possibly associated with flaring M dwarfs \citep{kow07}. The bimodality similar
to the one in the $g$ band is also discernible in the $r$ band, while it is less
pronounced in the $i$ band and not detected in the $u$ and $z$ bands (the $r$
and $z$ data are not shown).

\begin{figure}
\epsscale{1}
\plottwo{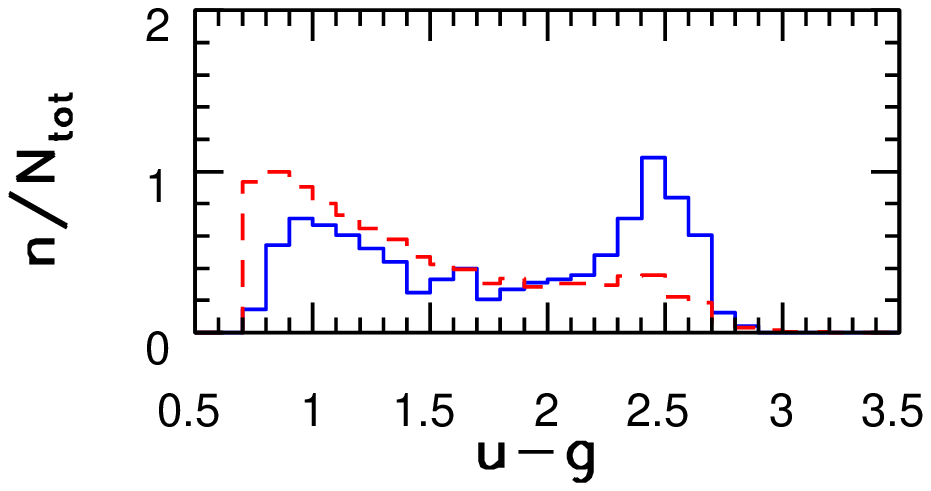}{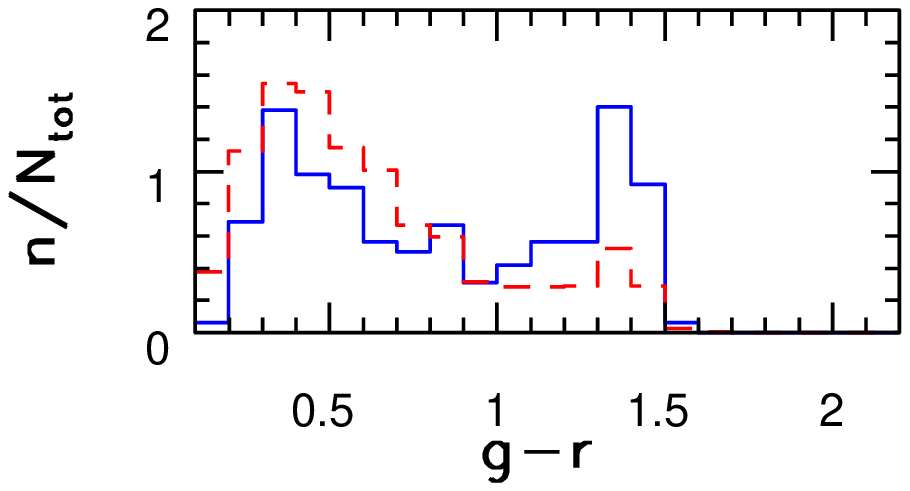}
\caption{
A comparison of the $u-g$ (left) and $g-r$ (right) color distributions for
variable main stellar locus stars brighter than $g=19$ (dashed lines), and a
subset with highly asymmetric lightcurves ($\gamma(g)>2.5$, solid lines). The
subset with highly asymmetric lightcurves has an increased fraction of stars
with colors $u-g\sim2.5$ and $g-r\sim1.5$, characteristic of M stars.
\label{skew_loc}}
\end{figure}

A comparison of the $u-g$ and $g-r$ color distributions for variable main
stellar locus stars brighter than $g=19$ and a subset with highly asymmetric
lightcurves ($\gamma(g)>2.5$) is shown in Figure~\ref{skew_loc}. The subset
with asymmetric lightcurves has an increased fraction of stars with colors
$u-g\sim2.5$ and $g-r\sim1.4$, that correspond to M stars. This may indicate
that M stars have a higher probability of being associated with an eclipsing
companion than stars with earlier spectral types. However, the selection effects
are probably important since a companion is easier to detect (due to the low
luminosity of M dwarfs). \citet{kow07} examine these issues using lightcurve
data on a sample of spectroscopically confirmed M dwarfs. Finally, quasars have
symmetric lightcurves ($\gamma\sim0$) and their distribution of skewness does
not change between bands.

\section{The Milky Way Halo Structure Traced by Candidate RR Lyrae Stars
\label{RRLyr}}

Studies of substructures in the Galactic halo, such as clumps and streams, can
constrain the formation history of the Milky Way. One of the best tracers to
study the outer halo are RR Lyrae stars because they are nearly standard
candles, are sufficiently bright to be detected at large distances ($5-100$ kpc
for $14<r<20.7$), and are sufficiently numerous to trace the halo substructure
with a high spatial resolution. The General Catalog of Variable Stars (GCVS;
\citealt{GCVS4}) lists\footnote{A list of GCVS variability types can be
found at
\url[HREF]{http://www.sai.msu.su/groups/cluster/gcvs/gcvs/iii/vartype.txt}}
RR Lyrae stars as RR Lyrae type $ab$ (RRab) and type $c$ (RRc) stars. RRab stars
have asymmetric lightcurves, periods from 0.3 to 1.2 days, and amplitudes from
$V\sim0.5$ to $V\sim2$. RRc stars have nearly symmetric, sometimes sinusoidal,
lightcurves, with periods from 0.2 to 0.5 days, and amplitudes not greater than
$V\sim0.8$. In this work we assume $M_V=0.7$ as the absolute $V$ band magnitude
of RRab and RRc stars. A comprehensive review of RR Lyrae stars can be found in
\citet{smi95}.

In this section we fine tune criteria for selecting candidate RR Lyrae stars,
and estimate the selection completeness and efficiency. Using selected candidate
RR Lyrae stars, we recover a known halo clump associated with the Sgr dwarf
tidal stream, and find several new halo substructures. 

\subsection{Criteria for Selecting RR Lyrae Stars\label{RR_selection}}

Figures~\ref{map_counts},~\ref{rmsg_col}, and~\ref{med_rms} show that RR
Lyrae stars occupy a well-defined region (Region IV) in the $g-r$ vs.~$u-g$
color-color diagram, and Figure~\ref{plot_Bayes} shows how RR Lyrae stars follow
the $\sigma(g)=1.4\sigma(r)$ relation. Motivated by these results, we introduce
color and $\sigma(g)/\sigma(r)$ cuts to specifically select candidate RR Lyrae
stars from the variable sample, and study their distribution in the
rms~scatter-color-lightcurve~skewness parameter space.

RR Lyrae stars have distinctive colors and can be selected with the following
criteria \citep{ive05}:
\begin{equation}
0.98 < u - g < 1.30
\label{color_cuts1}
\end{equation}
\begin{equation}
-0.05 < D_{ug} < 0.35
\end{equation}
\begin{equation}
0.06 < D_{gr} < 0.55
\end{equation}
\begin{equation}
-0.15 < r- i < 0.22
\end{equation}
\begin{equation}
-0.21 < i -z < 0.25
\end{equation}
where
\begin{equation}
D_{ug} = (u - g) +0.67(g-r)-1.07
\end{equation}
and
\begin{equation}
D_{gr} = 0.45(u-g) -(g-r)-0.12.
\label{color_cuts2}
\end{equation}

\begin{figure}
\epsscale{0.7}
\plotone{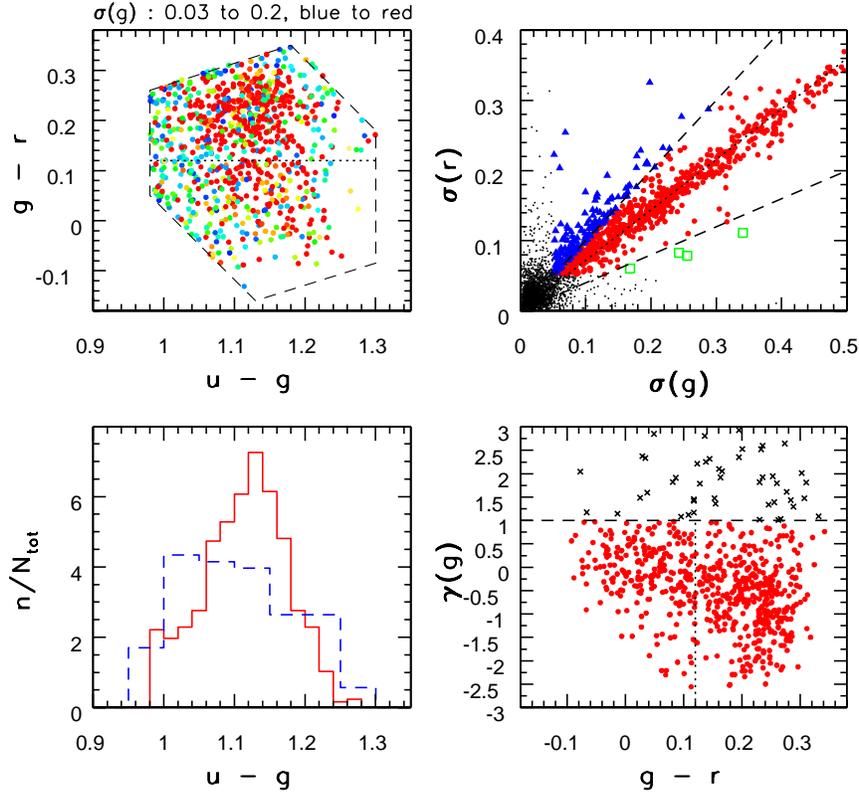}
\caption{
{\em Top left:} The distribution of 846 candidate variable sources from the
RR Lyrae region (dashed lines, see Fig.~3 in \citealt{ive05}) in the $g-r$
vs.~$u-g$ color-color diagram. The symbols mark the time-averaged values and
are color-coded by $\sigma(g)$ (0.05 to 0.2, blue to red). The dotted horizontal
line shows the boundary between the RRab and RRc-dominated regions.
{\em Top right:} Sources from the top left panel divided into 3 groups according
to their $\sigma(g)/\sigma(r)$ values: candidate RR Lyrae stars with
$1<\sigma(g)/\sigma(r)\leqslant 2.5$ (large dots), sources with
$\sigma(g)/\sigma(r)\leqslant1$ (triangles), and sources with
$\sigma(g)/\sigma(r)>2.5$ (squares). Small dots show sources with RR Lyrae
colors that fail the variability criteria. The dashed lines show the
$\sigma(g)=\sigma(r)$ and $\sigma(g)=2.5\sigma(r)$ relations, while the dotted
line shows the $\sigma(g)=1.4\sigma(r)$ relation. {\em Bottom left:} A
comparison of the $u-g$ color distributions for candidate RR Lyrae stars (solid
line) and sources with RR Lyrae colors but not tagged as RR Lyrae stars (dashed
line). {\em Bottom right:} The dependence of $\gamma(g)$ on the $g-r$ color for
candidate RR Lyrae stars. The boundary $g-r=0.12$ (vertical dotted line)
separates candidate RR Lyrae stars into those with asymmetric
($\gamma(g)\sim-0.5$) and symmetric ($\gamma(g)\sim0$) lightcurves,
corresponding to RRab and RRc stars, respectively. The condition
$\gamma(g)\leqslant1$ (horizontal dashed line) is used to reduce the
contamination of the RR Lyrae sample by eclipsing variables.
\label{RRBox}}
\end{figure}

We apply these cuts to our sample of candidate variables and select 846 sources.
It is implied by \citet{ive05} that RR Lyrae should always stay within these
color boundaries, even though their colors change as a function of phase.
Their distribution in the $g-r$ vs.~$u-g$ color-color diagram and rms scatter in
the $g$ band are shown in Figure~\ref{RRBox} (top left panel). The distribution
of sources in the RR Lyrae region is inhomogeneous. Sources with large rms
scatter in the $g$ band ($\ga0.2$ mag) are centered around $u-g\sim1.15$, and
are separated by $g-r\sim0.12$ into two groups. A comparison with Figure~3 from
\citet{ive05} suggests that these large rms scatter sources might be RR Lyrae
type $ab$ (RRab, $g-r>0.12$) and type $c$ stars (RRc, $g-r<0.12$). Small rms
scatter sources ($\la0.1$ mag) have a fairly uniform distribution, and are
slightly bluer with $u-g\la1.1$.

The distribution of sources from the RR Lyrae region in the $\sigma(r)$
vs.~$\sigma(g)$ diagram is presented in the top right panel of
Figure~\ref{RRBox}. The majority of large rms scatter sources follow the
$\sigma(g)=1.4\sigma(r)$ relation, as expected for RR Lyrae stars. Since RR
Lyrae stars are bluer when brighter, or equivalently, have greater rms scatter
in the $g$ band than in the $r$ band, we require
$1<\sigma(g)/\sigma(r)\leqslant2.5$ and select 683 candidate RR Lyrae stars.

A comparison of $u-g$ color distributions for candidate RR Lyrae stars and of
sources with RR Lyrae colors, but not tagged as RR Lyrae stars, presented in
the bottom left panel of Figure~\ref{RRBox}, demonstrates the robustness of
the RR Lyrae selection. The two distributions are very different (the
probability that they are the same is $10^{-4}$, as given by the KS test), with
the candidate RR Lyrae distribution peaking at $u-g\sim1.15$, as expected for RR
Lyrae stars.

One property that distinguishes RRab from RRc stars is the shape (or skewness)
of their lightcurves (in addition to lightcurve amplitude and period). RRab
stars have asymmetric lightcurves, while RRc lightcurves are symmetric. In the
top left panel of Figure~\ref{RRBox}, we noted that $g-r\sim0.12$ seemingly
separates high rms scatter sources into two groups. If $g-r\sim0.12$ is the
boundary between the RRab and RRc stars, then the same boundary should show up
in the distribution of lightcurve skewness as a function of the $g-r$ color. As
shown in Figure~\ref{RRBox} (bottom left panel), this is indeed the case. On
average, sources with $g-r<0.12$ have $\gamma(g)\sim0$ (symmetric lightcurves),
as RRc stars, while $g-r>0.12$ sources have $\gamma(g)\sim-0.5$ (asymmetric
lightcurves) typical of RRab stars.

We show in Section~\ref{RR_eff} that candidate RR Lyrae stars with
$\gamma(g)>1$ are contaminated by eclipsing variables. Therefore, to reduce the
contamination by eclipsing variables, we also require $\gamma(g)\leqslant1$, and
select 634 sources as our final sample of candidate RR Lyrae stars.

\subsection{Completeness and efficiency\label{RR_eff}}

The selection completeness, defined as the fraction of recovered RR Lyrae stars,
will depend on the color cuts, $\sigma(g,r)$ cutoff, and the number of
observations. The color cuts (Eqs.~\ref{color_cuts1} to~\ref{color_cuts2})
applied in Section~\ref{RR_selection} were chosen to minimize contamination
by sources other than RR Lyrae stars while maintaining an almost $100\%$
completeness \citep{ive05}. With the $\sigma(g,r)$ cutoff at 0.05 mag (small
compared to the $\sim1$ mag typical peak-to-peak amplitudes of RR Lyrae stars),
and a fairly large number of observations per source (median of 10), we estimate
the RR Lyrae selection completeness to be $\ga95\%$ (see Appendix in
\citealt{ive00}).

\begin{figure}
\epsscale{0.8}
\plotone{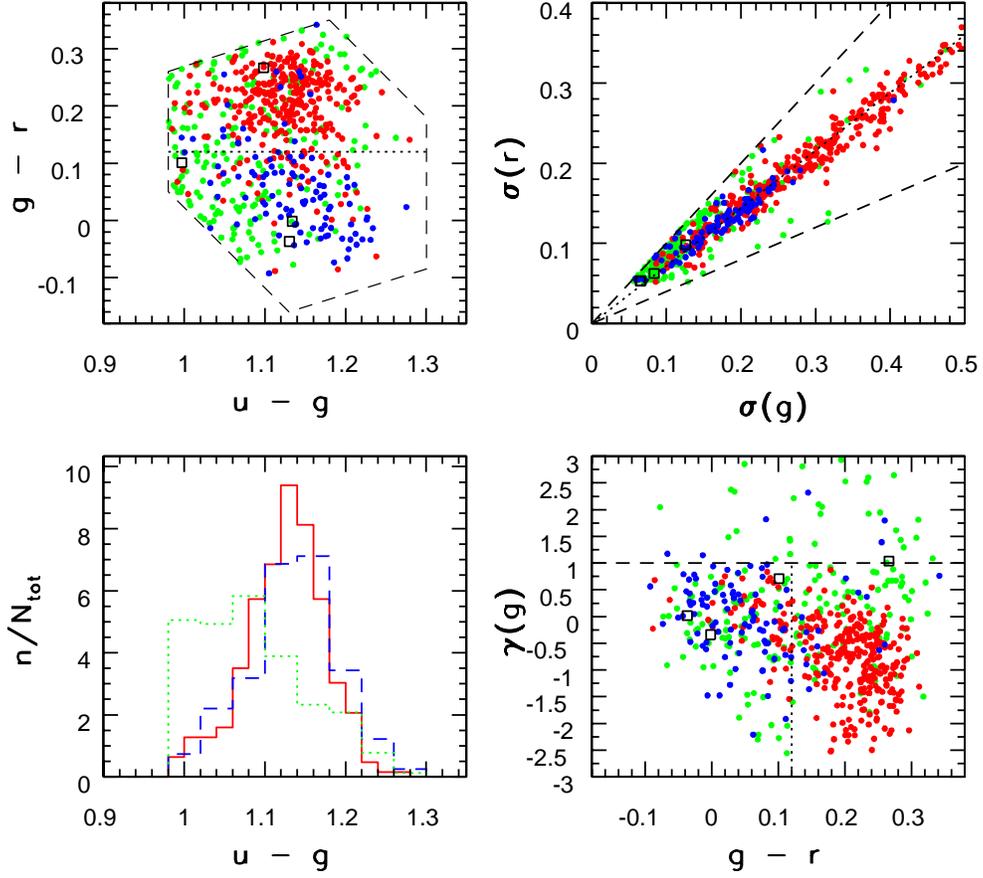}
\caption{
The distribution of candidate RR Lyrae stars selected with
$1<\sigma(g)/\sigma(r)\leqslant 2.5$ and classified by \citet{lee07}, shown in
diagrams similar to Fig.~\ref{RRBox}. Symbols show RRab stars (red dots), RRc
stars (blue dots), variable non-RR Lyrae stars (green dots), and non-variable
sources (open squares, only four sources). A comparison of the $u-g$ color
distribution for RRab (solid line), RRc (dashed line), and variable non-RR Lyrae
stars (dotted line) is shown in the bottom left panel.
\label{rrlyr_box2}}
\end{figure}

To determine the selection efficiency, defined as the fraction of true RR Lyrae
stars in the RR Lyrae candidate sample, we positionally match 683 candidate RR
Lyrae stars selected by $1<\sigma(g)/\sigma(r)\leqslant2.5$ to a sample of
RR Lyrae sources selected from the SDSS Light-Motion-Curve Catalog (LMCC;
\citealt{bra07}). This catalog covers the same region of the sky as the one
discussed here, but includes more recent SDSS-II observations that allow the
construction of lightcurves. We match 613 candidates, while 70 candidate RR
Lyrae stars from our sample, for some reason, do not have a match in the LMCC
(De Lee, private communication). Following the classification based on phased
lightcurves by \citet{lee07}, we find that $71\%$ of sources in our candidate RR
Lyrae sample are classified as RRab and RRc, $28\%$ are classified as variable
non-RR Lyrae stars, and only $1\%$ are spurious, non-variable sources. The most
significant contamination comes from a population of variable sources bluer than
$u-g\sim1.1$ (dotted line, bottom left panel Figure~\ref{rrlyr_box2}), possibly
Population II $\delta$ Scuti stars, also known as SX Phoenicis stars
\citep{hrw85}.

The top left and the bottom right panels in Figure~\ref{rrlyr_box2}, show that
RRab and RRc-dominated regions are separated by $g-r\sim0.12$, as already hinted
in Figure~\ref{RRBox}. Also, variable non-RR Lyrae sources with $\gamma(g)>1$
are classified by \citet{lee07} as eclipsing variables, justifying our
$\gamma(g)\leqslant1$ cut.

To summarize, using color criteria and criteria based on $\sigma(g)$,
$\sigma(r)$, and $\gamma(g)$ RR Lyrae stars are selected with $\ga95\%$
completeness and $\sim70\%$ efficiency.

\subsection{The Spatial Distribution of Candidate RR Lyrae Stars}

\begin{figure}
\epsscale{1.0}
\plotone{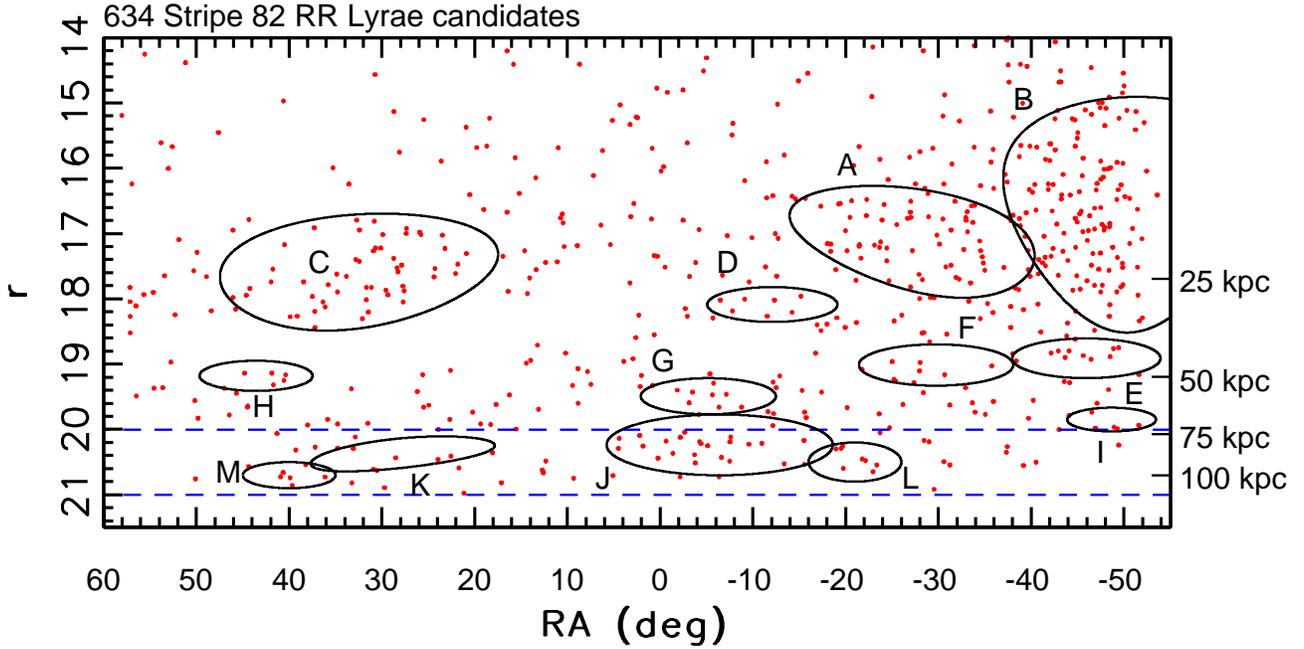}
\caption{
The magnitude-position distribution of 634 Stripe 82 RR Lyrae candidates within
$-55\arcdeg < R.A. < 60\arcdeg$ and $|Dec|\leqslant 1.27\arcdeg$. Approximate
distance (shown on the right y-axis) is calculated assuming $M_r = 0.7$ mag for
RR Lyrae stars. Dashed lines show where sample completeness decreases from
approximately $99\%$ to $60\%$ due to the $\chi^2$ cut (see the bottom right
panel in Fig.~\ref{chi_cum}). Closed curves are remapped ellipses and circles
from Fig.~\ref{rrlyr_polarslice} that mark halo substructure.
\label{RArEQ}}
\end{figure}

Using the selection criteria from Section~\ref{RR_selection} we isolate 634
RR Lyrae candidates. The magnitude-position diagram for these candidates
within $2.5\arcdeg$ from the Celestial Equator is shown in Figure~\ref{RArEQ}.

\begin{figure}
\epsscale{1.0}
\plottwo{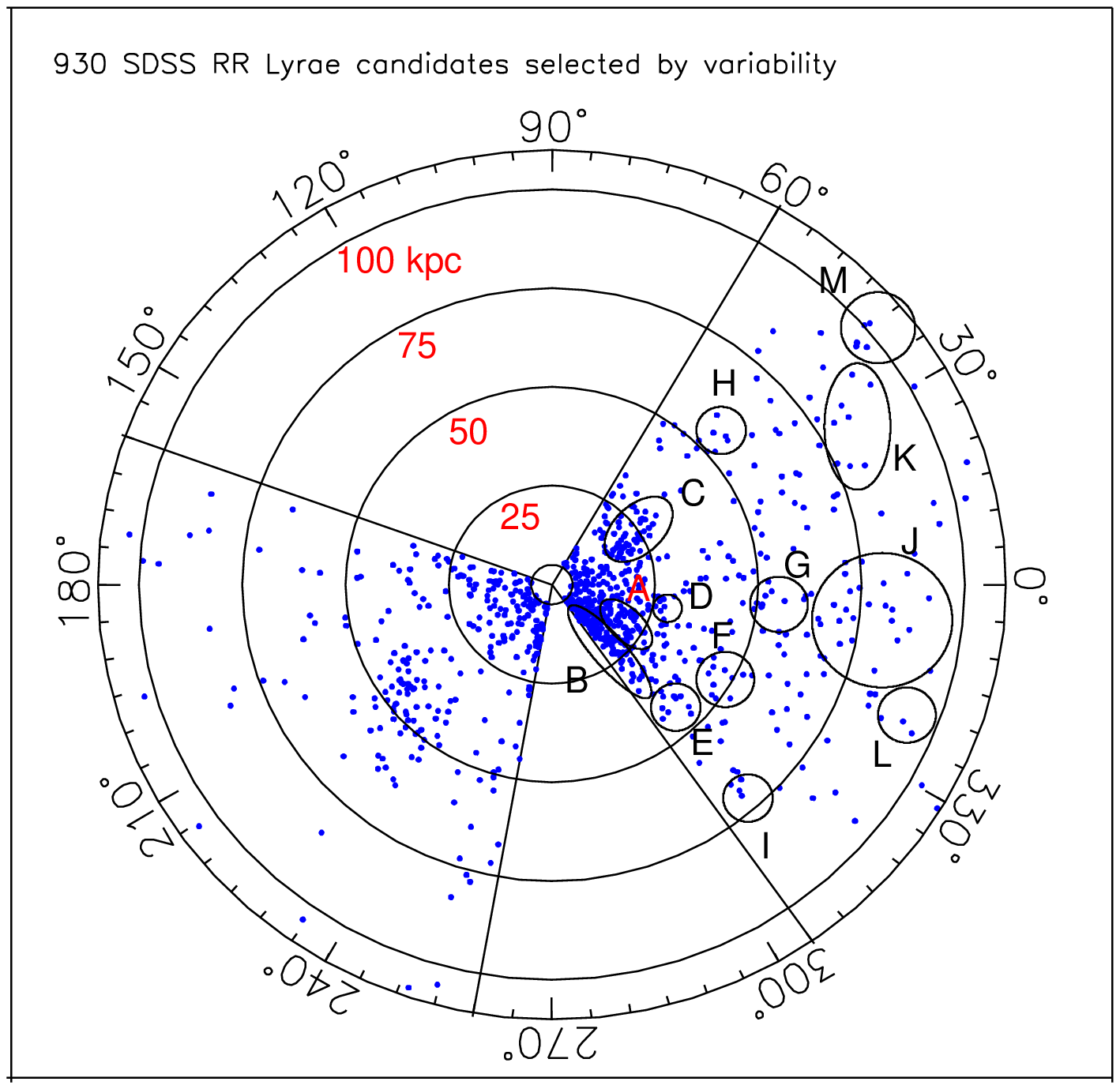}{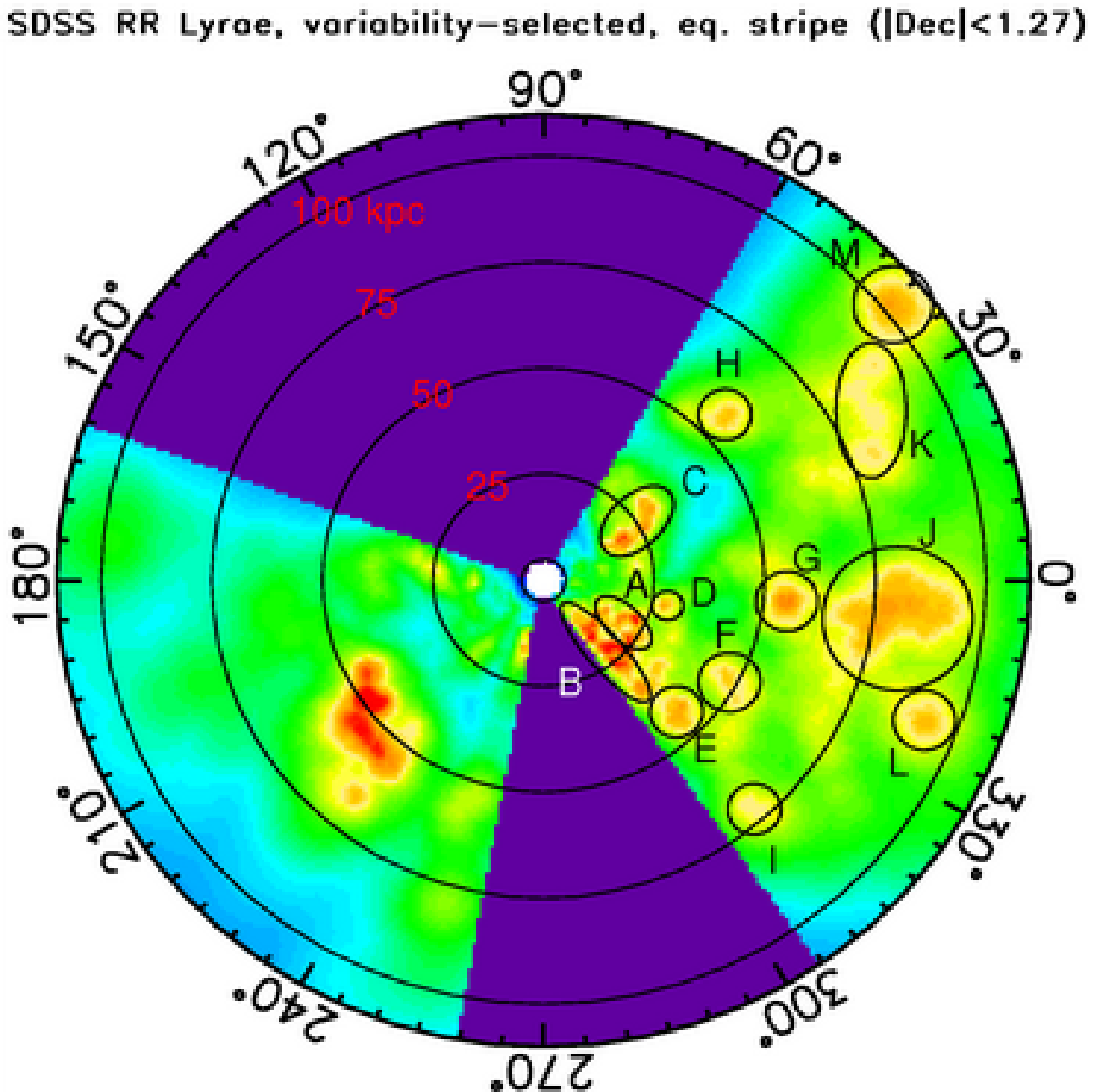}
\caption{
{\em Left:} The spatial distribution of candidate RR Lyrae stars discovered by
SDSS along the Celestial Equator. Distance is calculated assuming Eq. 3 from
\citet{ive05} and $M_V=0.7$ mag as the absolute magnitude of RR Lyrae in the
$V$ band. The right wedge corresponds to candidate RR Lyrae selected in this
work (634 candidates, shown in Fig.~\ref{RArEQ}) and the left wedge is based on
the sample from \citet{ive00} (296 candidates). {\em Right:} The number density
distribution of candidate RR Lyrae stars shown in the left panel, computed using
an adaptive Bayesian density estimator developed by \citet{ive05}. The color
scheme represents the number density multiplied by the cube of the
galactocentric radius, and displayed on a logarithmic scale with a dynamic range
of 300 (from light blue to red). The green color corresponds to the mean density
-- both wedges with the data would have this color if the halo number density
distribution followed a perfectly smooth $r^{-3}$ power-law. The purple color
marks the regions with no data. The yellow regions are formally $\sim3\sigma$
significant overdensities, and orange/red regions have an even higher
significance (using only the counts variance). The strongest clump in the left
wedge belongs to the Sgr dwarf tidal stream as does the clump marked by $C$ in
the right wedge \citep{ive03a}. An approximate location and properties of
labeled overdensities are listed in Table~\ref{halo}. The \citet{ive00} sample
is based on only 2 epochs and thus has a much lower completeness ($\sim56\%$)
resulting in a lower density contrast.
\label{rrlyr_polarslice}}
\end{figure}

As discussed by \citet{ive05}, an advantage of the data representation utilized
in Figure~\ref{RArEQ} (magnitude--right ascension diagram) is its simplicity --
only ``raw'' data are shown, without any post-processing. However, the magnitude
scale is logarithmic and thus the spatial extent of structures is heavily
distorted. In order to avoid these shortcomings, we have applied a Bayesian
method for estimating continuous spatial density distribution developed by
\citet{ive05} (see their Appendix B). The resulting density map is shown in the
right panel in Figure~\ref{rrlyr_polarslice}. The advantage of that
representation is that it better conveys the significance of various local
overdensities. For comparison, we also show a map of the northern part of
the equatorial strip constructed using 2-epoch data discussed by \citet{ive00}.

We detect several new halo substructures at $\ga3\sigma$ significance (compared
to expected Poissonian fluctuations) and present their approximate locations and
properties in Table~\ref{halo}. The most distant clump is at 100 kpc from the
Galactic center. The strongest clump in the left wedge belongs to the Sgr dwarf
tidal stream as does the clump marked by $C$ in the right wedge \citep{ive03a}.
We note that the apparent ``clumpiness'' of the candidate RR Lyrae distribution
increases with increasing radius, similar to CDM predictions by \citet{bkw01}. A
detailed comparison of their models with the data presented here will be
discussed elsewhere (Sesar et al., in prep).

\section{Are All Quasars Variable?\label{QSO}}

The optical continuum variability of quasars has been recognized since their
first optical identification \citep{ms63}, and it has been proposed and utilized
as an efficient method for their discovery \citep{bhp73,haw83,kkc86,hv95}. The
observed characteristics of the variability of quasars are frequently used to
constrain the origin of their emission (e.g. \citealt{kaw98} and references
therein; \citealt{ms03,per06}). Recently, significant progress in the description of
quasar variability has been made by employing the SDSS data
\citep{dVBW03,ive04b,VB04,dV05,ses06}. Here we expand these studies by
quantifying the efficiency of quasar discovery using variability.

A preliminary comparison of color and variability based methods for selecting
quasars using SDSS data was presented by \citet{ive03b}. They found that $47\%$
of spectroscopically confirmed unresolved quasars with UV excess have the $g$
band magnitude difference between two observations obtained two years apart
larger than 0.15 mag. We can improve on their analysis because now there are
significantly more observations obtained over a longer time period. Since
quasars vary erratically and the rms scatter of their variability (the so-called
structure function) increases with time (e.g. \citealt{VB04} and references
therein), the variability selection completeness is expected to be higher than
$\sim50\%$ obtained by \citet{ive03b}. 

First, although the adopted variability selection criteria discussed above are
fairly conservative, we find that at least $63\%$ of low-redshift quasars are 
variable at the $\geqslant0.05$ mag level (simultaneously in the $g$ and $r$
bands over observer's time scales of several years) in the $g<20.5$ flux-limited
sample. Second, even this estimate is only a lower limit: given the
spectroscopic confirmation for a large flux-limited sample of quasars, it is
possible to relax the adopted variability selection cutoff without a prohibitive
contamination by non-variable sources. 

There are 2,492 unresolved quasars in the catalog of spectroscopically confirmed
SDSS quasars \citep{sch05} from Stripe 82. The fraction of these objects that
vary more than $\sigma$ in the $g$ and $r$ bands, as a function of $\sigma$, is
shown in Figure~\ref{qso_rms}. We also show the analogous fraction for stars
from the stellar locus. About $93\%$ of quasars vary with $\sigma>0.03$ mag. For
a small fraction of these objects the measured rms scatter is due to
photometric noise, and the stellar data limit this fraction to be at most $3\%$.
Conservatively assuming that none of these $3\%$ of stars is intrinsically
variable, we estimate that {\em at least $90\%$ of quasars are variable at the
0.03 mag level on time scales up to several years}.

\section{Implications for Surveys such as LSST\label{LSSTsec}}

The Large Synoptic Survey Telescope (LSST) is a proposed imaging survey that
aims to obtain repeated multi-band imaging to faint limiting magnitudes over a
large fraction of the sky. The LSST Science Requirement Document\footnote{
Available at \url[HREF]{http://www.lsst.org/Science/lsst\_baseline.shtml}} calls
for $\sim1000$ repeated observations of a solid angle of $\sim20,000$ deg$^2$
distributed over the six $ugrizY$ photometric bandpasses and over 10 years. The
results presented here can be extrapolated to estimate the lower limit on the
number of variable sources that the LSST would discover. 

The single-epoch LSST images will have a $5\sigma$ detection limit\footnote{An
LSST Exposure Time Calculator is available at \url[HREF]{www.lsst.org}} at
$r\sim24.7$. Hence, $2\%$ accurate photometry, comparable to the subsample with
$g<20.5$ discussed here, will be available for stars with $r\la22$. The USNO-B
catalog \citep{mon03} shows that there are about $10^9$ stars with $r<21$ across
the entire sky. About half of these stars are in the parts of the sky to be
surveyed by the LSST. The simulations based on contemporary Milky Way models,
such as those developed by \citet{rob03} and \citet{jur07}, predict that there
are about twice as many stars with $r<22$ than with $r<21$ across the whole sky.
Hence, it is expected that the LSST will detect about a billion stars with
$r<22$. This estimate is uncertain to within a factor of two or so due to
unknown details in the spatial distribution of dust in the Galactic plane and
towards the Galactic center. 

We found that at least $0.5\%$ of stars from the main stellar locus can be
detected as variable with photometry accurate to $\sim2\%$. This is only a lower
limit because a much larger number of LSST observations obtained over a longer
timespan than the SDSS data discussed here would increase this fraction. Hence,
our results imply that {\it the LSST will discover at least 50 million variable
stars} (without accounting for the fact that stellar counts greatly increase
closer to the Galactic plane). Unlike the SDSS sample, where RR Lyrae stars
account for $\sim25\%$ of all variable {\it stars}, the number of RR Lyrae stars
in the LSST sample will be negligible compared to other types of variable stars.

As estimated by \citet{jur07} using deeper coadded SDSS photometry, there are 
about 100 deg$^{-2}$ low-redshift quasars with $r<22$ (see also \citealt{bwa07}
and references therein). Therefore, with a sky coverage of $\sim20,000$ deg$^2$,
the LSST will obtain well-sampled accurate multi-color lightcurves for $\sim2$
million low-redshift quasars. Even at the redshift limit of $\sim2$, this sample
will be complete to $M_r \sim -24$, that is, almost to the formal quasar
luminosity cutoff, and will represent an unprecedented sample for studying
quasar physics.

\section{Conclusions and Discussion\label{discussion}}

We have designed and tested algorithms for selecting candidate variable sources 
from a catalog based on multiple SDSS imaging observations. Using a sample
of 13,051 selected candidate variable sources in the adopted $g<20.5$ 
flux-limited sample, we find that at least $2\%$ of unresolved optical sources 
appear variable at the $\geqslant0.05$ mag level, simultaneously in the $g$ and 
$r$ bands. A similar fraction of variable sources ($\sim1\%$) was also found by 
\citet{ses06} using recalibrated photometric POSS and SDSS measurements, and 
by \citet{mor06} using the Faint Sky Variability Survey data ($\sim1\%$).

Thanks to the multi-color nature of the SDSS photometry, and especially to the
$u$ band data, we can obtain robust classification of selected variable sources.
The majority (2/3) of variable sources are low-redshift ($<2$) quasars, although
they represent only $2\%$ of all sources in the adopted $g<20.5$ flux-limited
sample. We find that about 1/4 of variable stars are RR Lyrae stars, and that
only $0.5\%$ of stars from the main stellar locus are variable at the 0.05 mag
level. 

The distribution of $\gamma(g)$ for main stellar locus stars is bimodal,
suggesting at least two, and perhaps more, different populations of variables.
About a third of variable stars from the stellar locus show gray flux variations
in the $g$ and $r$ bands ($\sigma(g)/\sigma(r)\sim1$), and positive lightcurve
skewness, suggesting variability caused by eclipsing systems. This population
has an increased fraction of M type stars.

RR Lyrae stars show the largest rms scatter in the $u$ and $g$ bands, followed
by low-redshift quasars. The ratio of rms scatter in the $g$ and $r$ bands for
RR Lyrae is $\sim1.4$, in agreement with \citet{ive00} results based on 2-epoch
photometry. The mean lightcurve skewness for RR Lyrae stars is $\sim-0.5$, in
agreement with \citet{wlb06}. We selected a sample of 634 candidate RR Lyrae
stars, with an estimated $\ga95\%$ completeness and $\sim70\%$ efficiency. Using
these stars, we detected rich halo substructure out to distances of 100 kpc. The
apparent ``clumpiness'' of the candidate RR Lyrae distribution increases with 
increasing radius, similar to CDM predictions by \citet{bkw01}. 

\begin{figure}
\epsscale{1.0}
\plotone{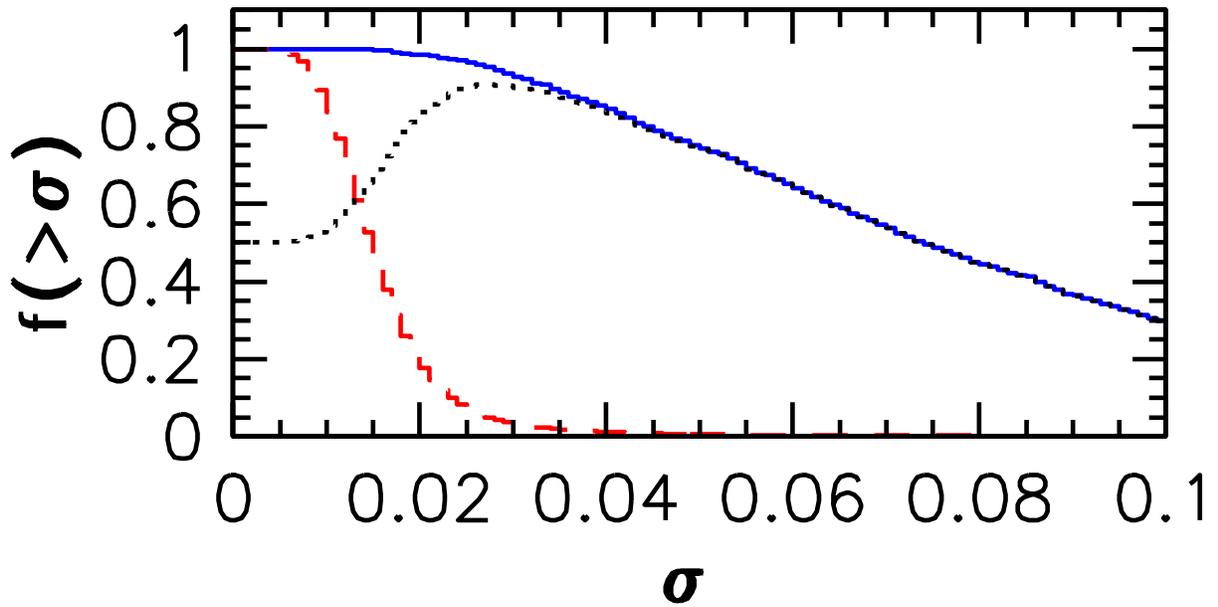}
\caption{
The fraction of spectroscopically confirmed unresolved QSOs ($f_{QSO}$, solid
line) and the fraction of sources from the stellar locus ($f_{loc}$, dashed
line) brighter than $g=19.5$ and $r=19.5$ that have rms scatter larger than
$\sigma$ in the $g$ and $r$ bands. The ratio $f_{QSO}/(1+f_{loc})$ (dotted
line), which corresponds to the implied fraction of variable QSOs, peaks at a
level of $90\%$ for $\sigma=0.03$ mag. 
\label{qso_rms}}
\end{figure}

Low-redshift quasars show a dependence of $\sigma(g)/\sigma(r)$ on redshift,
consistent with discussions in \citet{ric02} and \citet{wil06}. The lightcurve
skewness distribution for quasars is centered on zero in all photometric bands.
We find that at least $90\%$ of quasars are variable at the 0.03 mag level (rms)
on time scales up to several years. This confirms that variability is as a
good a method for finding low-redshift quasars at high ($|b|>30$) Galactic
latitudes as is the UV excess color selection. The fraction of variable quasars
at the $\geqslant0.1$ mag level obtained here ($30\%$, see Figure~\ref{qso_rms})
is comparable to 36\% found by \citet{rbw06}.

The multiple photometric observations obtained by the SDSS represent an
excellent dataset for estimating the impact of surveys such as the LSST on
studies of the variable sky. Our results indicate that the LSST will obtain
well-sampled $2\%$ accurate multi-color lightcurves for $\sim2$ million
low-redshift quasars, and will discover at least 50 million variable stars. The
number of variable stars discovered by the LSST will be of the same order as the
number of {\it all} stars detected by the SDSS. With about 1000 data points in
six photometric bands, it will be possible to recognize and classify variable
objects using lightcurve moments of higher order than skewness discussed here,
including lightcurve folding for periodic variables.

\acknowledgments

We acknowledge support by NSF grant AST-0551161 to the LSST for design and
development activity. 

Funding for the SDSS and SDSS-II has been provided by the Alfred P. Sloan
Foundation, the Participating Institutions, the National Science Foundation, the
U.S. Department of Energy, the National Aeronautics and Space Administration,
the Japanese Monbukagakusho, the Max Planck Society, and the Higher Education
Funding Council for England. The SDSS Web Site is http://www.sdss.org/.

The SDSS is managed by the Astrophysical Research Consortium for the
Participating Institutions. The Participating Institutions are the American
Museum of Natural History, Astrophysical Institute Potsdam, University of Basel,
University of Cambridge, Case Western Reserve University, University of Chicago,
Drexel University, Fermilab, the Institute for Advanced Study, the Japan
Participation Group, Johns Hopkins University, the Joint Institute for Nuclear
Astrophysics, the Kavli Institute for Particle Astrophysics and Cosmology, the
Korean Scientist Group, the Chinese Academy of Sciences (LAMOST), Los Alamos
National Laboratory, the Max-Planck-Institute for Astronomy (MPIA), the
Max-Planck-Institute for Astrophysics (MPA), New Mexico State University, Ohio
State University, University of Pittsburgh, University of Portsmouth, Princeton
University, the United States Naval Observatory, and the University of
Washington.

\clearpage


\begin{deluxetable}{rlrrrrrrrrrrrr}
\rotate
\tabletypesize{\scriptsize}
\tablecolumns{14}
\tablewidth{0pc}
\tablecaption{The distribution of candidate variable sources in the $g-r$ vs $u-g$ diagram\label{regions}}
\tablehead{
\multicolumn{2}{c}{} & \multicolumn{4}{c}{$g < 19$} & \multicolumn{4}{c}{$g < 20.5$} & \multicolumn{4}{c}{$g < 22$} \\
\cline{1-14} \\
\colhead{Region$^a$} & \colhead{Name$^b$} & 
\colhead{\% all$^c$} & \colhead{\% var$^d$} & \colhead{var/all$^e$} & \colhead{$N_{var}/N_{all}^f$} & 
\colhead{\% all$^c$} & \colhead{\% var$^d$} & \colhead{var/all$^e$} & \colhead{$N_{var}/N_{all}^f$} &
\colhead{\% all$^c$} & \colhead{\% var$^d$} & \colhead{var/all$^e$} & \colhead{$N_{var}/N_{all}^f$}
}
\startdata
I   & white dwarfs & 0.14    & 0.59  & 4.25  & 3.50  & 0.24    & 0.40   & 1.69  & 3.34  & 0.28    & 0.45   & 1.64  & 4.51   \\
II  & low-redshift QSOs   & 0.45    & 30.88 & 68.83 & 56.58 & 1.90    & 62.90  & 33.03 & 65.10 & 4.07    & 70.01  & 17.22 & 47.30  \\
III & dM/WD pairs & 0.08    & 0.53  & 6.54  & 5.37  & 0.83    & 2.08   & 2.50  & 4.92  & 1.21    & 3.79   & 3.13  & 8.61  \\
IV  & RR Lyrae stars   & 1.28    & 16.81 & 13.11 & 10.78 & 1.33    & 7.95   & 5.99  & 11.81 & 1.48    & 6.41   & 4.33  & 11.90  \\
V   & stellar locus stars  & 96.27   & 48.77 & 0.51  & 0.42  & 94.49   & 25.15  & 0.27  & 0.52  & 91.89   & 18.33  & 0.20  & 0.55  \\
VI  & high-redshift QSOs  & 1.78    & 2.42  & 1.36  & 1.12  & 1.21    & 1.52   & 1.26  & 2.48  & 1.07    & 1.01   & 0.95  & 2.60  \\
\tableline
    & total count  & 411,667 & 3,384 &       &       & 662,195 & 13,051 &       &       & 748,067 & 20,553 &       & 
\enddata
\tablenotetext{a}{These regions are defined in the $g-r$ vs. $u-g$ color-color
diagram, with their boundaries shown in Fig.~\ref{rmsg_col}}
\tablenotetext{b}{An approximate description of the dominant source type found
in the region} 
\tablenotetext{c}{The fraction of all sources in a magnitude-limited sample
found in this color region, with the magnitude limits listed on top.} 
\tablenotetext{d}{The number of candidate variables from the region, expressed
as a fraction of all variable sources}
\tablenotetext{e}{The ratio of values listed in columns d) and c)}
\tablenotetext{f}{The number of candidate variables from the region, expressed
as a fraction of all sources in that region}
\end{deluxetable}

\clearpage

\begin{deluxetable}{crrrr}
\rotate
\tabletypesize{\scriptsize}
\tablecolumns{9}
\tablewidth{0pc}
\tablecaption{The fraction of variable main stellar locus stars as a function of the $s$ color\label{principal}}
\tablehead{
\colhead{Bin} & \colhead{\% $\sigma(g)\geqslant0.05^a$} & \colhead{\% var$^b$} &
\colhead{$\langle \sigma(g)\rangle^c$} & \colhead{Counts$^d$} 
}
\startdata
$s<-0.02$      & 3.23  & 0.36 & 0.017 & 46,836  \\
$-0.02<s<0.02$ & 2.92  & 0.28 & 0.017 & 136,910 \\
$0.02<s<0.06$  & 4.61  & 1.18 & 0.019 & 29,106  \\
$s>0.06$       & 11.50 & 4.10 & 0.027 & 4,547
\enddata
\tablenotetext{a}{Fraction of sources with $\sigma(g)\geqslant0.05$ mag}
\tablenotetext{b}{Fraction of variable sources (selected using $\sigma(g,r)\geqslant0.05$ mag and $\chi^2(g,r)\geqslant3$)}
\tablenotetext{c}{Median $\sigma(g)$} 
\tablenotetext{d}{Number of sources in the bin} 
\end{deluxetable}

\clearpage

\begin{deluxetable}{rrrrrrrrr}
\rotate
\tabletypesize{\scriptsize}
\tablecolumns{9}
\tablewidth{0pc}
\tablecaption{Approximate locations and properties of detected overdensities\label{halo}}
\tablehead{
\colhead{Label$^a$} & \colhead{N$^b$} & \colhead{$\langle R.A.\rangle^c$} &
\colhead{$\langle d\rangle^d$} & \colhead{$\langle r\rangle^e$} &
\colhead{$\langle u-g\rangle^f$} & \colhead{$\langle g-r^g\rangle$} & 
\colhead{$\langle \gamma(g)\rangle^h$} & \colhead{$N_b/N_r^i$}
}
\startdata
A & 84  & 330.95 & 21  & 17.02 & 1.14 & 0.18 & -0.50 & 0.62 \\
B & 144 & 309.47 & 22  & 16.76 & 1.12 & 0.16 & -0.57 & 0.64 \\
C & 54  & 33.69  & 25  & 17.61 & 1.13 & 0.20 & -0.68 & 0.29 \\
D & 8   & 347.91 & 29  & 18.02 & 1.14 & 0.23 & -0.50 & 0.38 \\
E & 11  & 314.06 & 43  & 18.84 & 1.09 & 0.20 & -0.41 & 0.75 \\
F & 11  & 330.26 & 48  & 19.16 & 1.07 & 0.20 & -0.46 & 0.38 \\
G & 10  & 354.81 & 55  & 19.46 & 1.10 & 0.22 & -0.69 & 0.38 \\
H & 7   & 43.57  & 57  & 19.32 & 1.05 & 0.04 &  0.06 & 1.34 \\
I & 4   & 311.34 & 72  & 19.98 & 1.08 & 0.11 & -0.10 & 2.0  \\
J & 26  & 353.58 & 81  & 20.21 & 1.11 & 0.20 & -0.27 & 0.58 \\
K & 8   & 28.39  & 84  & 20.35 & 1.10 & 0.20 &  0.14 & 0.44 \\
L & 3   & 339.01 & 92  & 20.45 & 1.06 & 0.16 &  0.08 & 0.67 \\
M & 5   & 39.45  & 102 & 20.73 & 1.07 & 0.11 &  0.36 & 1.67
\enddata
\tablenotetext{a}{Overdensity's label from Fig.~\ref{rrlyr_polarslice}}
\tablenotetext{b}{Number of candidate RR Lyrae in the overdensity} 
\tablenotetext{c}{Median Right Ascension} 
\tablenotetext{d}{Median distance (in kpc)}
\tablenotetext{e}{Median $r$ band magnitude}
\tablenotetext{f}{Median $u-g$ color}
\tablenotetext{g}{Median $g-r$ color}
\tablenotetext{h}{Median $\gamma(g)$}
\tablenotetext{i}{The number ratio of candidate RR Lyrae with $g-r<0.12$ and $g-r>0.12$}
\end{deluxetable}

\clearpage


\clearpage

\clearpage

\clearpage

\clearpage

\clearpage

\clearpage

\clearpage

\clearpage

\clearpage

\clearpage

\clearpage

\clearpage

\clearpage

\end{document}